\documentclass[bibyear]{aa} 

\usepackage{graphicx}
\usepackage{txfonts}
\begin{document}

\title{The 2.3 GHz continuum survey of the GEM project\thanks{The survey is only available at the CDS via anonymous ftp to cdsarc.u-strasbg.fr (130.79.128.5) 
or via http://cdsweb.u-strasbg.fr/cgi-bin/qcat?J/A+A/}}

\author{C.~Tello\inst{1} \and T.~Villela\inst{1} \and S.~Torres\inst{2}
	\and M.~Bersanelli\inst{3} \and G.~F.~Smoot\inst{4,5} \and I.~S.~Ferreira\inst{6}
        \and A.~Cingoz\inst{5} \and J.~Lamb\inst{7} \and D.~Barbosa\inst{8}
        \and D.~Perez-Becker\inst{5,9} \and S.~Ricciardi\inst{4,10} \and J.~A.~Currivan\inst{5}
      	\and P.~Platania\inst{11} \and D.~Maino\inst{3}
          }


    \institute{Divis\~ao de Astrof\'\i sica,
        Instituto Nacional de Pesquisas Espaciais (INPE), CP 515, 12201-970, S\~ao Jos\'e dos Campos, SP, Brazil
    \and
        Centro Internacional de F\'\i sica, Bogot\'a, Colombia
    \and
    	Dipartimento di Fisica, Universit\`a degli Studi di Milano, via Celoria 16, 20133 Milano, Italy
    \and
        Lawrence Berkeley National Laboratory, University of California, 1 Cyclotron Road, Bldg. 50, MS 205, Berkeley, CA 94720, USA
    \and
        Physics Department, University of California, Berkeley CA 94720, USA
    \and
	Instituto de F\'\i sica, Universidade de Bras\'\i lia, Campus Universit\'ario Darcy Ribeiro - Asa Norte, 70919-970, Bras\'\i lia, DF, Brazil
    \and
        Physics Department, University of California, Santa Barbara CA 93106, USA
    \and
        Grupo de RadioAstronomia, Instituto de Telecomunica\c c\~oes, Campus Universit\'ario de Aveiro, Aveiro, Portugal
    \and
        Departamento de F\'\i sica, Facultad de Ciencias, Universidad Nacional Aut\'onoma de M\'exico, 04510 DF, Mexico
    \and
        INAF Osservatorio Astronomico di Padova, Vicolo dell'Osservatorio 5, I - 35122, Padova, Italy
    \and
        Istituto di Fisica del Plasma, CNR-ENEA-EURATOM Association, Via R. Cozzi 53, 20125 Milano, Italy
             }

   \date{Received 1 March 2013 ; Accepted 19 April 2013 }


  \abstract
{Determining the spectral and spatial characteristics of the radio continuum of our Galaxy is an experimentally challenging endeavour for improving our 
understanding of the astrophysics of the interstellar medium. This knowledge has also become of paramount significance for cosmology, since Galactic emission 
is the main source of astrophysical contamination in measurements of the cosmic microwave background (CMB) radiation on large angular scales.}
   {We present a partial-sky survey of the radio continuum at 2.3\thinspace{GHz} within the scope of the Galactic Emission Mapping (GEM) project,
an observational program conceived and developed to reveal the large-scale properties of Galactic synchrotron radiation through a set of self-consistent surveys of the radio continuum between 408\thinspace{MHz}\ and 10\thinspace{GHz}.}
   {The GEM experiment uses a portable and double-shielded 5.5-m radiotelescope in altazimuthal configuration to map 60-degree-wide declination bands
from different observational sites by circularly scanning the sky at zenithal angles of 30$\degr$ from a constantly rotating platform. The observations were
accomplished with a total power receiver, whose front-end High Electron Mobility Transistor (HEMT) amplifier was matched directly to a cylindrical horn at the 
prime focus of the parabolic reflector. The Moon was used to calibrate the antenna temperature scale and the preparation of the map required direct subtraction
and destriping algorithms to remove ground contamination as the most significant source of systematic error.}
   {We used 484 hours of total intensity observations from two locations in Colombia and Brazil to yield 66\% sky coverage from $\delta = -51\fdg73$ to
$\delta = +34\fdg78$. The observations in Colombia were obtained with a horizontal HPBW of $2\fdg30\pm0\fdg13$ and a vertical HPBW of $1\fdg92\pm0\fdg18$. 
The pointing accuracy was $6\farcm84$ and the RMS sensitivity was $11.42$ mK. The observations in Brazil were obtained with a horizontal HPBW of $2.31\pm0.03$ 
and a vertical HPBW of $1\fdg82\pm0\fdg12$. The pointing accuracy was $5\farcm26$ and the RMS sensitivity was $8.24$ mK. The zero-level uncertainty of the combined 
survey is 103\thinspace{mK} with a temperature scale error of 5\% after direct correlation with the Rhodes/HartRAO survey at 2326\thinspace{MHz} on a $T$-$T$ 
plot.}
   {The sky brightness distribution into regions of low and high emission in the GEM survey is consistent with the appearance of a transition region as seen
in the Haslam 408\thinspace{MHz} and WMAP K-band surveys. Preliminary results also show that the temperature spectral index between 408\thinspace{MHz} and the 
2.3\thinspace{GHz} band of the GEM survey has a weak spatial correlation with these regions; but it steepens significantly from high to low emission regions 
with respect to the WMAP K-band survey.}

   \keywords{surveys -- Galaxy: structure -- radio continuum: general -- radio continuum: ISM}

   \titlerunning{The GEM 2.3 GHz survey}
   \authorrunning{Tello et al.}

   \maketitle
%

\section{Introduction}
The large-scale distribution of non-thermal radiation from our Galaxy has been observed since the birth of radioastronomy at very low frequencies
of a few tens of MHz (Jansky \cite{Jansky32}; \cite{Jansky33}; \cite{Jansky35}) until the advent of the present era of precision cosmology at frequencies above 
20\thinspace{GHz} (e.g. Mather et al. 1990; Smoot et al. 1992; de Bernardis et al. 2000; Bennett et al.~\cite{Bennett03}; Meinhold et al. 2005). 
By the time its synchrotron nature was recognized to be due to relativistic electrons spiralling along interstellar magnetic field 
lines (Kiepenheuer \cite{Kiepen50}; Mayer et al.~\cite{MMS57}; Mills \cite{Mills59}; Westerhout \cite{Wester62}; 
Wielebinski et al.~\cite{Wiel62}), free-free radiation from HII regions had already been identified as another Galactic emission component through 
the pioneering work of Reber (\cite{Reber40}), who sketched the first radio maps of the Galaxy at 160\thinspace{MHz} and 480\thinspace{MHz} 
(Reber \cite{Reber44}; Reber \cite{Reber48}). Excluding atmospheric emission and the 3\thinspace{K} due to the cosmic microwave background (CMB), the sky signal on large angular scales is dominated by synchrotron radiation up to $\sim$20\thinspace{GHz} and by free-free and anomalous microwave emission (AME) up to $\sim$70\thinspace{GHz}, 
where the rising spectrum of thermal dust takes over the characterization of Galactic emission all the way into the infrared, including rotational line emission of CO molecules above 100\thinspace{GHz} (Planck Collaboration XII \cite{PlanckXII-13}; Planck Collaboration XIII \cite{PlanckXIII-13}). Given this association 
with distinct astrophysical sources, each Galactic emission component traces a unique spatial template. In the case of the non-thermal continuum,
radioastronomers have always been actively pursuing the frequency dependance of the synchrotron template to understand the structure
and composition of the interstellar medium. In cosmology, the nature of temperature anisotropies in the CMB and its polarization cannot be
fully understood unless the contaminating role played by diffuse Galactic emission is accurately detailed.

In this article, we present the first survey of an observational program, the Galactic Emission Mapping (GEM) project, aimed at supplementing the
frequency gap in radio-continuum surveys that map its large-scale distribution at centimeter wavelengths and the all-sky maps 
from the {\it COBE}, {\it WMAP}, and {\it Planck} satellite missions in the millimetric and submillimetric bands. We explore this deficiency in Sect.~\ref{RCS} as we review the present status of the GEM project toward the preparation of spatial templates of the 
synchrotron component between 408\thinspace{MHz} and 10\thinspace{GHz}. In Sect.~\ref{EXP} we describe our portable single-dish scanning 
experiment at 2.3\thinspace{GHz}\ during three observing seasons in Colombia and Brazil. In Sect.~\ref{DDaP} we outline the data reduction and 
preparation of a combined map with 66\% of sky coverage in total intensity between $\delta = -51\fdg73$ and $\delta = +34\fdg78$. Its baseline
calibration and destriping techniques are treated in Sect.~\ref{MAP} together with the absolute temperature scale of the final map.
We discuss our results in the light of the Rhodes/HartRAO survey at 2326\thinspace{MHz} (Jonas et al.~\cite{Jonas98}) in Sect.~\ref{RES} and, finally, 
we state our conclusions and future prospects in Sect.~\ref{ENDE}.


\section{Large-scale Galactic emission and the GEM project}\label{RCS}
Unparalleled efforts have always been dedicated to delineate the complex morphology of the synchrotron-dominated Galactic emission at meter and
centimeter wavelengths, which reveals the rich environment of supernova remnants with spurs and loops emanating from the Galactic Plane
and, in some cases, engulfing our location near the inner Perseus spiral arm of the Galaxy. The work of Jansky and Reber led naturally to the
completion of the first all-sky radio surveys. The earliest of these was the 200\thinspace{MHz} survey of Dr\"oge and Priester (\cite{Droge56}), which
combined the observations of Allen and Gum (\cite{Allen50}) in the Southern Hemisphere. The advent of the world's largest radiotelescopes,
together with an improvement in the performance of the receivers, prompted the realization of several partial surveys of better sensitivity and
resolution at 178\thinspace{MHz} (Turtle \& Baldwin \cite{TB62}), 30\thinspace{MHz} (Mathewson et al.~\cite{MS65}), 38\thinspace{MHz} (Milogradov-Turin
\& Smith \cite{M73}), 85\thinspace{MHz} and 150\thinspace{MHz} (Yates et al.~\cite{Y67}; Landecker \& Wielebinski \cite{LW70}), which were used to
synthesize three additional low-frequency all-sky surveys: the 30\thinspace{MHz} map of Cane (\cite{Cane78}), the 85\thinspace{MHz} map of
Yates (\cite{Y68}), and the 150\thinspace{MHz} map of Landecker \& Wielebinski (\cite{LW70}). The integrity of these maps is, however,
compromised by the need to assume spectral indices when combining different frequencies and by significant sidelobe level corrections due to
differential ground and sky pick-up; let alone the differences in the mapping strategy of each radiotelescope.

A major improvement in the preparation of all-sky surveys was achieved by Haslam et al.~(\cite{Haslam81}; \cite{Haslam82}), whose 
408\thinspace{MHz} map of the whole sky reflects the choice of a single observing frequency and mapping strategy across four surveys to obtain a 
resolution under $1\degr$ with negligible sidelobe contamination using single-dish observations. More recently, another all-sky medium-resolution
($\sim 0\fdg5$) survey was completed at 1420\thinspace{MHz} through the observations of Reich \& Reich (\cite{Reich82}; \cite{Reich86}) in the Northern 
Hemisphere and those of Testori et al.~(\cite{Testo01}) in the Southern Hemisphere (Reich et al.~\cite{Reich01}). Single-dish surveys of the radio 
continuum for mapping the large-scale Galactic emission have not been pursued at higher frequencies, with the exception of the Rhodes/HartRAO 
survey at 2326 MHz (Jonas et al.~\cite{Jonas98}) over 67\% of the sky with a HPBW = $0\fdg333$. Higher resolution surveys in this frequency range 
have been obtained with single-dish antennas at 1.4\thinspace{MHz} (Reich et al.~\cite{Reich90a}; Reich et al.~\cite{Reich97}), 2.4\thinspace{MHz} 
(Duncan et al.~\cite{Duncan95}), and 2.7\thinspace{MHz} (Reich et al.~\cite{Reich84}; Reich et al.~\cite{Reich90b}; F\"urst et al.~\cite{Fuerst90}); 
but their sky coverage has been limited to within a few degrees of the Galactic Plane. At lower frequencies, low-resolution antenna arrays have been 
used to conclude three other major surveys with 69\% and 95\% of total sky coverage, respectively, at 22\thinspace{MHz} (Roger et al.~\cite{Roger99}) 
and 45\thinspace{MHz} (Alvarez et al.~\cite{Alvar97}; Maeda et al.~\cite{Maeda99}). 

{In astrophysics, medium- and low-resolution surveys complement high-resolution surveys in probing the full range of spatial scales that are necessary 
to explore the complexity of the ISM. In particular, toward the Galactic disk, where the majority of the Galactic radio sources and the 21-cm 
continuum emission of atomic hydrogen are concentrated.  Combining dedicated single-dish observations and their sensitivity to diffuse emission 
with interferometric observations down to the resolution limit in the Galactic disk has resulted in the International Galactic Plane Survey of HI, 
which covers 90\% of the Galactic disk extending across three surveys: the Canadian Galactic Plane Survey, or CGPS (Taylor et al.~\cite{Taylo03}), 
the Southern Galactic Plane Survey, or SGPS (McClure-Griffiths et al.~\cite{McClu05}), and the VLA Galactic Plane Survey, or VGPS 
(Stil et al.~\cite{Stil06}).

\subsection{CMB contamination}\label{CMB}
The detection of temperature anisotropies in the CMB radiation by the DMR experiment onboard
the {\it COBE} satellite (Smoot et al.~\cite{Smoot92}) initiated a scientific breakthrough for cosmology; but it also showed, from the analysis of the
{\it COBE} all-sky maps at 30\thinspace{GHz}, 53\thinspace{GHz}, and 90\thinspace{GHz} (Bennett et al.~\cite{Bennett92}), the need to extend our knowledge of
diffuse Galactic emission to higher frequencies. Experiments to measure the CMB with higher precision and resolution have since then been developed or are 
currently under way. These new generations of experiments are conducted from the ground (e.g. BEAST -- Childers et al.~\cite{Childers05}; Meinhold et
al.~\cite{Meinhold05}; O'Dwyer et al.~\cite{O'Dwyer05}; QUIET -- QUIET Collaboration \cite{QUIET11}; \cite{QUIET12}; SPT -- Keisler at al.~\cite{Keisler11}; Reichardt et al.~\cite{Reicha12}; ACT -- Das et al.~\cite{Das11}; D\"unner et al.~\cite{Dunner13}), from balloon platforms (e.g. BOOMERanG -- de Bernardis et al.~\cite{Bernardis00}; EBEX -- Reichborn-Kjennerud et al.~\cite{Reichb10}), or onboard satellites (WMAP -- Bennett et al.~\cite{Bennett03}; Gold et al.~\cite{Gold11}; Planck -- Tauber et al.~\cite{Tau10}; Planck Collaboration I \cite{PlanckI-11}). All these efforts have established polarization measurements as one of the hottest topics in observational cosmology. The success of high-precision 
CMB observations imposes an even more reliable and accurate separation of the cosmological signal from the foreground emission of our own 
Galaxy, which cannot be removed by any improvement in the detectors, whether on the ground or in space. Several techniques have become 
available; from simple Galaxy masking of heavily contaminated regions and point sources (e.g. Mej\'\i a et al.~\cite{Mejia05}) to the novel approaches 
of component separation and foreground removal (Hobson et al.~\cite{Hobson98}; Maino et al.~\cite{Maino02}; Delabrouille et 
al.~\cite{Delabr03}; \cite{Delabr09}; Eriksen et al.~\cite{Eriksen04}; \cite{Eriksen06}; \cite{Eriksen08}; Bonaldi et al.~\cite{Bonaldi06}; Cardoso et al.~\cite{Cardo08}; Leach et al.~\cite{Leach08}; Stompor et al.~\cite{Stom09}; Ricciardi et al.~\cite{Ricciardi10}; Fern\'andez-Cobos et al.~\cite{Fernan12}; Planck Collaboration XII \cite{PlanckXII-13}), 
which exploit the spectral and spatial diversity of the different 
astrophysical and cosmological components. The multi-frequency design of the {\it WMAP} mission had already revealed an impressive picture of 
unprecedented precision of the CMB at 23\thinspace{GHz}, 33\thinspace{GHz}, 41\thinspace{GHz}, 61\thinspace{GHz} and 94\thinspace{GHz}
(Hinshaw et al.~\cite{Hinshaw07}; \cite{Hinshaw12}; Bennett et al.~\cite{Bennett12}); whereas the {\it Planck} mission, with its higher angular resolution and superb sensitivity at 30\thinspace{GHz}, 44\thinspace{GHz}, 70\thinspace{GHz}, 100\thinspace{GHz}, 143\thinspace{GHz}, 217\thinspace{GHz}, 353\thinspace{GHz}, 545\thinspace{GHz}, and 857\thinspace{GHz} has, at the time of writing, made available an overwhelming set of results that push the standards of cosmological and astrophysical research to a state-of-the-art level (Planck Collaboration I \cite{PlanckI-13}).

Yet, there remains a considerable gap in our knowledge of diffuse Galactic emission between the upper-frequency end of the large 
ground-based radio-telescopes and the lower end achievable from space or balloons, which confronts radioastronomers and cosmologists alike. 
In this frequency range, the increased resolution of single-dish observations imply that the time allocation at radio-astronomical facilities capable 
of surveying a sizable area of the sky, meaningful enough to address the large-scale properties of this diffuse component, becomes practically 
prohibitive. The few examples we find in the literature, where high-resolution surveys were undertaken in this frequency gap, default to efforts 
devoted to construct catalogues of radio sources against the high-intensity emission of the Galactic disk. Among them, we find arcminute-resolution 
surveys at 4.9\thinspace{GHz} (Altenhoff et al.~\cite{Alten79}), 5\thinspace{GHz} (Haynes et al.~\cite{Haynes78}), and 10\thinspace{GHz} 
(Handa et al.~\cite{Handa87}). Without a dedicated instrument, free of time constraints in its observational schedule, diffuse Galactic emission will
continue to lack the coverage it needs to map the sky away from the Galactic Plane in this frequency gap. This is a crucial step in our understanding
of the astrophysics of the ISM with a direct impact for the cosmological problem of foreground contamination. First of all, without a detailed 
morphological description of the variation of the synchrotron spectral index across the sky, the usefulness of a synchrotron spatial template looses its 
appeal in the study of CMB anisotropies. Second, the roll-off in the energy spectrum of relativistic electrons, as they leave their confinement volumes
in the Galactic disk, is expected to introduce breaks in the spectrum of the non-thermal component up to a few GHz, with significant implications
regarding (a) the elusive nature of anomalous microwave emission (AME) from spinning dust grains (Draine \& Lazarian \cite{DraLaz98}; \cite{DraLaz99}; 
Lagache \cite{Lagache03}; de Oliveira-Costa et al.~\cite{Angel04}; Watson et al.~\cite{Watson05}; Davies et al.~\cite{Davies06}; Dobler and Finkbeiner \cite{Dobler08}; Dickinson et al.~\cite{Dick09}; Ysard et al.~\cite{Ysard10}; Planck Collaboration XX \cite{PlanckXX-11}); (b) 
the identification of spectral distortions in the black body curve of the CMB signal (Burigana et al.~\cite{Burigana91}; Bartlett \& Stebbins 
\cite{Bartlett91}); and more recently, (c) the residual radio signal in the ARCADE data (Seiffert et al.~\cite{Seiffert11}; 
Kogut et al.~\cite{Kogut11}; Fixsen et al.~\cite{Fixsen11}) and (d) the existence of the Galactic haze (Finkbeiner \cite{Fink04}; Dobler \cite{Dobler12}; Planck Collaboration Int. IX \cite{PlanckIX-12}).

The lack of surveys in this frequency gap is even more critical for polarization surveys. Synchrotron radiation is intrinsically polarized and its
contaminating role poses higher risks for polarization than for total intensity. In addition, the selection effect of Faraday depolarization is partial
only toward the lower end of the gap. The first all-sky absolutely-calibrated polarization survey (Wolleben et al.~\cite{Wolle06}; Testori et al.~\cite{Testo04}) was a rather recent achievement at 1.4\thinspace{GHz}. Still, the gap extends up to the lowest of the WMAP frequency
bands (Page et al.~\cite{Page07}; Hinshaw et al.~\cite{Hinshaw09}). From the point of view of cosmological implications, it becomes more urgent to bridge 
radio and microwave polarization surveys; whereas from the astrophysical point of view, the precise elucidation of the Galaxy as a Faraday screen is of 
key importance to the study of magnetic fields in the Galaxy (Beck \cite{Beck07}) via the synchrotron mechanism. Clearly, polarization surveys at 
frequencies higher than a few GHz (for example Leonardi et al.~\cite{Leonardi06}) are needed to address the contamination of the CMB at higher 
frequencies, where the current Planck mission, with a wider frequency coverage and a better sensitivity, will undoubtedly deepen our cosmological 
understanding (Fauvet et al.~\cite{Fauvet12}). Despite their lack of sensitivity to diffuse Galactic emission, Galactic Plane surveys also produce a wealth of information regarding the
magneto-ionic medium, which help us understand the complexity of the interplay between electron densities and magnetic field strengths in generating 
the pervading synchrotron radiation of the low-frequency sky (Gao et al.~\cite{Gao10}). Below we outline the scope of a novel observational program 
aimed to bridge the mapping of diffuse Galactic emission in this frequency gap.

\subsection{The Galactic Emission Mapping project}\label{GEM}
The concept of the GEM project dates back to 1992 when the first detections of CMB anisotropies called for an improved understanding of diffuse
Galactic emission (Smoot et al.~\cite{Smoot92}; Bennett et al.~\cite{Bennett92}) to eliminate foreground contamination from genuine
cosmic signals. The existing radio-continuum surveys suffered from two major shortcomings. First, they lacked mutual consistency in their 
baseline levels (Lawson et al.~\cite{Lawson87}; Banday \& Wolfendale \cite{Banday91}) to be able to reliably predict the spectrum of
Galactic synchrotron emission in the cosmologically interesting windows. Second, the spatial templates traced by the non-thermal component in the 
408\thinspace{MHz} and 1420\thinspace{MHz} surveys were compromised by striping along the scanning direction of the
radiotelescopes (Davies et al.~\cite{DWG96}). Thus, given the frequency gap mentioned above, the initial goal of the GEM project became 
the production of an atlas of total sky brightness maps at 408\thinspace{MHz}, 1465\thinspace{MHz}, 2.3\thinspace{GHz}, 5\thinspace{GHz} and 
10\thinspace{GHz}, which would overcome these undesired radioastronomical shortcomings (De Amici et al.~\cite{Amici94}; Smoot \cite{Smoot99})
using a full-time dedicated experiment from different observational sites. Improved mapping of the spectral and large-scale properties 
of Galactic synchrotron radiation would result from an observational strategy aimed to minimize ground contamination and long-term systematics to
guarantee baseline uniformity. In addition, the production of a self-consistent set of sky maps would allow the re-examination of the baseline and 
destriping in previous surveys to take advantage of their superior resolutions.

At the core of the GEM project stood the development of a portable 5.5-m double-shielded radiotelescope, which could be deployed at different
geographic latitudes to maximize homogeneous sky coverage. Since ground contamination is one of the most common sources of systematic errors in
this type of survey experiments, the two shields were built to minimize sidelobe pick-up of stray radiation and to level out the horizon profile at
the site location. These shields are schematically drawn in Fig.~\ref{Fig1} and consisted, respectively, of: (a) a 2.1-m-long rim-halo of 
aluminum panels extending tangentially from each of the 24 dish petals; and (b) a concentric network of inclined wire-mesh ground screens, which 
levels the horizon with a $10\degr$-high azimuth profile. Under full illumination, the dish subtends an angle of $158\degr$ at the prime focus. Tello 
et al.~(\cite{Tello99}; \cite{Tello00}) modeled the diffraction and attenuation properties of this double-shielded radiotelescope to estimate the 
spillover and diffraction sidelobe contamination seen in elevation scans with back-fire helical feeds at 408\thinspace{MHz}\ and 
1465\thinspace{MHz}.

\begin{figure}
\resizebox{8.8cm}{!}{\includegraphics{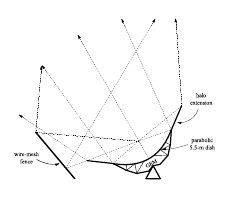}}
\caption{Ray-tracing diagram (dotted lines) of the illumination properties of the double-shielded GEM dish.}
\label{Fig1}
\end{figure}

Diurnal and seasonal variations in the temperature of the ground also add low-frequency noise to survey measurements carried out continuously
over extended periods of time. Similarly, the variability of atmospheric emission introduces, in addition to low-frequency noise, short-term
contributions on the scale of minutes. To make individual scans insensitive to atmospheric fluctuations while still being able to map large
areas of the sky, we adopted azimuthal scans of sufficiently high constant speed (1\thinspace{rpm} for $Z=30\degr$ scans at 2.3\thinspace{GHz}). 
Coupled to the rotation of the Earth, these zenith-centered circular scans achieve coverage of full declination bands with around-the-clock 
observations as shown schematically in Fig.~\ref{Fig2}, where sample circular scans, spaced at one-hour intervals, illustrate the 
coverage in declination for a site in the Southern Hemisphere.

\begin{figure}
\centering
\resizebox{8cm}{!}{\includegraphics{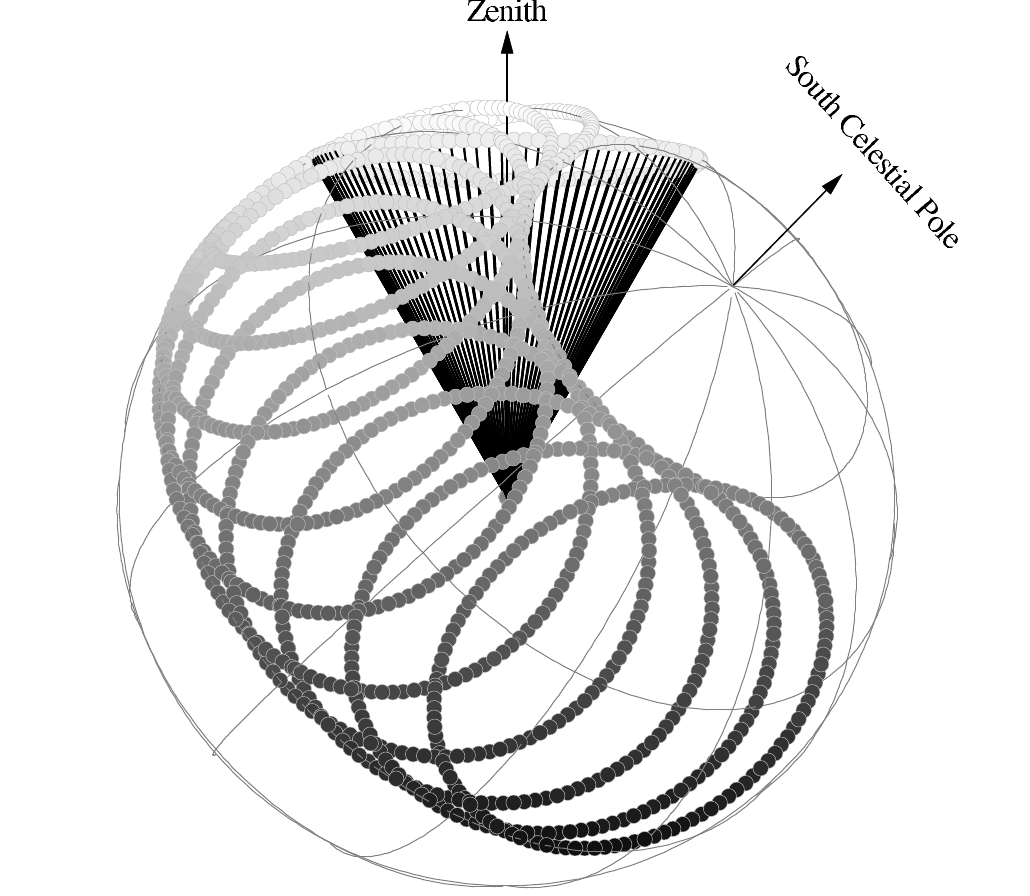}}
\caption{Schematic representation of the mapping strategy with the GEM experiment.}
\label{Fig2}
\end{figure}

\begin{table*}[ht]
\centering
\caption[]{Total intensity observations with the GEM experiment}
\begin{tabular}{rccrr}
\noalign{\vskip -6pt}
\hline\hline
\noalign{\vskip 2.5pt}
\multicolumn{1}{c}{Frequency}&\multicolumn{1}{c}{HPBW}&\multicolumn{1}{c}{Sky coverage}&\multicolumn{1}{c}{Duration}&\multicolumn{1}{r}{Observational campaign}\\
\multicolumn{1}{c}{(MHz)}&\multicolumn{1}{c}{(degrees)}&\multicolumn{1}{c}{(degrees)}&\multicolumn{1}{c}{(hours)}\\
\noalign{\vskip 2pt}
\hline
\noalign{\vskip 2.5pt}
408 & $10.5$ & $+07\fdg2 < \delta < +67\fdg5$ & 40 & Bishop, CA, USA 1994 (1)\\
1465 & $4.2$ & $+07\fdg2 < \delta < +67\fdg5$ & 70 & Bishop, CA, USA 1994 (2)\\
408 & $11.3$ & $-24\fdg4 < \delta < +35\fdg6$ & 447 & Villa de Leyva, Colombia 1995 (3)\\
2300 & $3.7$ & $-24\fdg4 < \delta < +35\fdg6$ & 231 & Villa de Leyva, Colombia 1995 (4)\\
1465 & $4.2$ & $-52\fdg1 < \delta < +06\fdg7$ & 709 & Cachoeira Paulista, SP, Brazil 1999 (5)\\
2300 & $2\fdg30\times 1\fdg85$ & $-51\fdg73 < \delta < +06\fdg36$ & 532 & Cachoeira Paulista, SP, Brazil 1999 (6)\\
1465 & $4.2$ & $-52\fdg1 < \delta < +06\fdg7$ & 510 & Cachoeira Paulista, SP, Brazil 2004--2005 (7)\\
1465 & $4.2$ & $-52\fdg1 < \delta < +06\fdg7$ & 417 & Cachoeira Paulista, SP, Brazil 2009 (7)\\
\noalign{\vskip 2.5pt}
\hline						
\noalign{\vskip 2.5pt}
\multicolumn{5}{l}{References: (1) Rodrigues \cite{Rogerio00}; (2) Tello \cite{Tello97}; (3) Torres et al.~\cite{Torres96}; 
(4) Torres et al.,~internal report;}\\
\multicolumn{5}{l}{(5) Tello et al.~\cite{Tello05}; (6) Tello et al.,~this article; (7) Tello et al.,~in preparation}\\
\noalign{\vskip 2.5pt}
\end{tabular}
\label{Tab1}
\end{table*}

Table \ref{Tab1} lists a summary of the main GEM observational campaigns since its first deployment in the Owens Valley desert near Bishop, CA,
in 1993. In this period, the ground- and rim-halo shields have been continuously refurbished and upgraded to achieve satisfactory levels of ground
rejection up to 10\thinspace{GHz}. A new rim-halo was re-designed and installed in 2001, whereas the ground screen has gone through two major 
upgrades to incorporate a finer mesh along with a more robust support structure (two ground shields have been lost due to unusually gusty and strong 
winds). Another great impediment for radio astronomical surveys is the ever increasing usage of the spectrum by human-related activities, which plagues
the radio environment with non-negligible levels of radio-frequency interference (RFI). 100\% duty-cycle interferers have aborted GEM operations
several times in the past (the 408\thinspace{MHz}\ campaign in 1994, the 1465\thinspace{MHz}\ runs in 1995, and follow-up observations at 
2.3\thinspace{GHz}\ in 2004). 

Not included in Table \ref{Tab1} are the polarization measurements we have been conducting at 5\thinspace{GHz} since 2006, which have already been used 
to prepare a preliminary version of our first polarization survey (Ferreira et al.~\cite{Ferreira08}) from the Southern Hemisphere. The recent inclusion 
of Portugal in the GEM collaboration promises to extend our 5\thinspace{GHz}\ and 10\thinspace{GHz}\ polarization templates of Stokes $I$, $Q$ and $U$ 
parameters to the Northern Hemisphere, which can provide substantial ancilliary data for the Planck satellite mission and other future CMB polarization 
experiments. The GEM-P program will certainly benefit from a carefully chosen site (Fonseca et al.~\cite{Fonseca06}).

\section{The experiment at 2.3\thinspace{GHz}}\label{EXP}
The portability of the GEM experiment reduces its operation to three main modules: (a) a dish pedestal with an altazimuthally rotating platform, (b) a
radiometer, and (c) a control unit. These are graphically displayed in Fig.~\ref{Fig3} with a slip-rings assembly aligned with the axis of rotation along
the base column of the dish pedestal. Their task is to distribute power to the scanning dish and to establish a communication protocol with the
radiometer. Two interface boxes, one inside a ground enclosure and another attached to the co-rotating base column, provide the connecting ports for the slip-rings.
A 1-hp AC-motor, controlled by frequency inverters and coupled to the main gear of the pedestal by means of a speed reducer, provided the
motion in azimuth. An optical shaft encoder was fixed to the bottom of the slip-rings assembly to measure the angular variations of the rotating
platform. Similarly, an elevation encoder was installed on the dish horizontal axis. The analog signals of both encoders are also sent to connecting
ports in their respective interface boxes, from where they are routed to the data acquisition box located on the elevation support arm beneath the dish.
The data acquisition system (DAS) assembles the data frame by digitizing the analog inputs from the encoders, the radiometer and a noise source,
before sending it seamlessly through the slip-rings up to the control panel for storage on a PC.

\begin{figure*}
\centering\includegraphics[width=15cm]{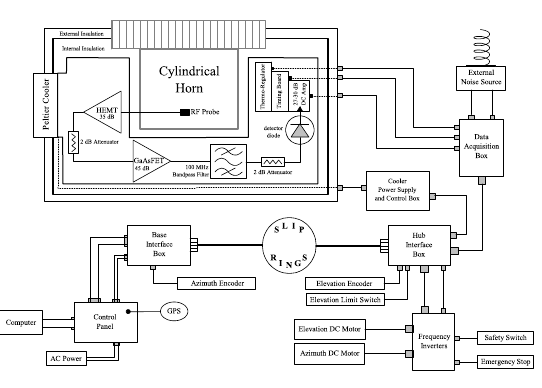}
\caption{Graphical layout of the GEM experiment at 2.3\thinspace{GHz}.}
\label{Fig3}
\end{figure*}

\subsection{Data acquisition system}\label{DAS}
The data acquisition system (DAS) uses three single-width NIM modules to interface with the experiment and a data-logging PC. The control module
has an EPROM with a 16-channel data frame burnt-in and an output connector for a serial output line. The analog multiplexer/ADC module receives
16 differential analog inputs of FSD $\pm10\mathrm{V}$ with a common-mode rejection of $86$\thinspace{dB}. The ADC limits hard at $\pm$10\thinspace{V}
with no bleed-over between channels and overloads of $\pm$30\thinspace{V}. In addition to the EOF signal, the DAS also times the firing of a thermally
stable noise source every 80 frames. The noise source is kept underneath the dish surface inside an insulated box and transmits an
$\approx$140\thinspace{K} reference signal from a small front-fire helix sitting at the junction of two of the dish petals. Two analog channels are
dedicated to record the temperature and the voltage of the noise source. Finally, a serial-to-RS232 module takes the seamless serial data and
synchronizes each frame with the hexadecimal EB90 word for decoding in the PC. LabView routines were used with Macintosh and
Pentium PC to log the data. Proper time-stamping of the data followed by adding UT from a WWV digital receiver clock to the data frame for the
observations in Colombia; whereas the WGS84 datum from a GPS receiver provided the time reference for the observations in Brazil.

\subsection{Radiometric characterization}\label{RCH}
For the observations at 2.3\thinspace{GHz}\ we used the three-legged support structure of the parabolic reflector to secure a direct-gain total-power 
radiometer at its focal plane by means of a latching mechanism. Its feed is built-in at the bottom of the radiometer enclosure, as shown in 
Fig.~\ref{Fig4}, and consists of a flanged cylindrical horn with $1/4$-$\lambda$ chokes and a measured voltage standing wave ratio (VSWR) of 1.023 at 2.3\thinspace{GHz}\ 
(return loss of $-39$\thinspace{dB}). Its circular aperture provides suitable illumination of the primary with a symmetric beam and a discrimination 
of $-16$\thinspace{dB} at the edge of the reflector. Since the Moon is a reasonable approximation of a point source for our beam, horizontal and 
vertical beam patterns were obtained directly from the survey data by mapping the signal onto a Moon-centered coordinate system. With our circular 
scanning strategy, the horizontal pattern is derived from the antenna response when the Moon reaches the same elevation as the scanning circle, 
while the vertical pattern is established from the antenna response when the dish azimuth coincides with that of the Moon. These beam pattern 
profiles are shown in Fig.~\ref{Fig3ini} for the data collected at the Colombian and Brazilian sites. Using Gaussian fits to these profiles, we 
obtained weighted average beam widths of $2\fdg31\pm0\fdg03$ for the horizontal and $1\fdg85\pm0\fdg10$ for the vertical planes of the pattern 
from the two experimental set-ups. Assuming that the secondary introduces no beam distortion, the 25\% larger horizontal beam is an upper limit to 
the smearing of the beam due to the rotation of the antenna. Using internal errors only, this smearing of the beam is significant at the $4.5\sigma$ level.

\begin{figure}
\resizebox{8.8cm}{!}{\includegraphics{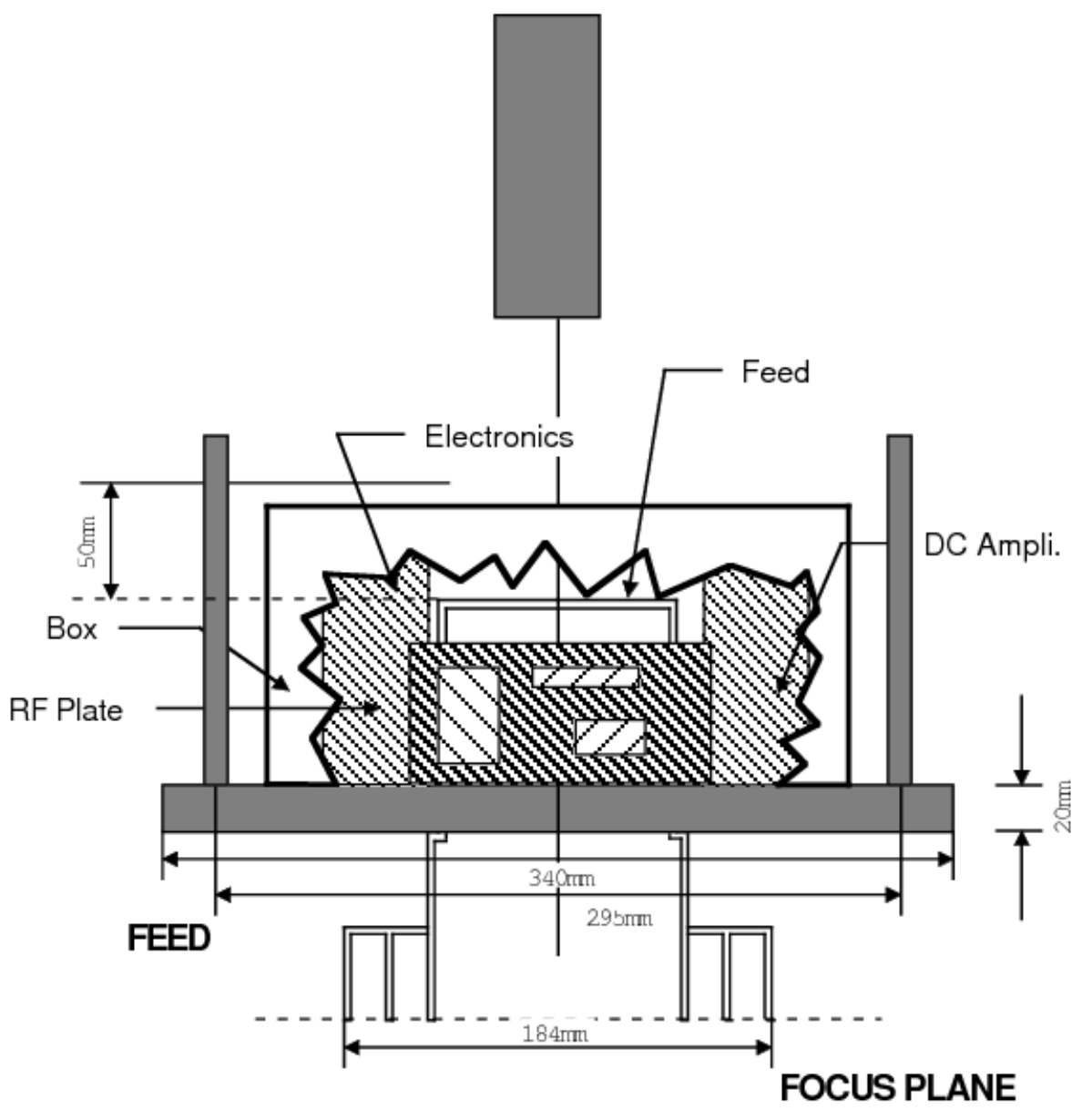}}
\caption{Focal configuration at 2.3\thinspace{GHz}\ with a direct-gain total power radiometer.}
\label{Fig4}
\end{figure}

The power fed by the reflector is captured at the throat of the horn by a waveguide probe, which sends it down a section of semi-rigid coax into a
front-end high electron mobility transistor (HEMT) amplifier. As depicted in Fig.~\ref{Fig3}, the RF chain of the receiver section attains a
nominal total amplification of $\sim$70\thinspace{dB}. The HEMT amplifier has a flat response across the entire 100\thinspace{MHz} width of the 
bandpass-defining filter and a noise temperature of $\approx$30\thinspace{K} at an operating temperature of $300$\thinspace{K}. After the filter, the signal is rectified by a square-law detector diode, and its DC voltage is amplified by a factor of 500 (low gain) or 1000 (high gain). The DC amplifier is housed with a
thermo-regulating circuit and a timing board inside a detached electronics box. The timing board subdivides the pulses from an end-of-data frame
(EOF) signal generated in the data acquisition box to sample the DC voltage with a time constant of $\tau=0.56002\mathrm{s}$.

For the receiver to achieve an optimal performance, its gain variations were kept to a minimum by thermally stabilizing the interior of the RF-tight
radiometer enclosure $\approx$10\thinspace{\degr C} below the temperature set-point of the aluminum RF plate. To accomplish this functionality, we
used a Peltier cooling unit on the side of the radiometer to continuously extract heat from its interior; while the thermo-regulating circuit maintained
the RF plate (physical) temperature at its set-point by means of a sensing diode on the plate to actively control a set of power resistors. The rest of the RF
enclosure was coated internally and externally with insulating styrofoam. Four calibrated temperature sensors constantly monitored the thermal
state of the radiometer at the detector diode, the HEMT amplifier, the horn flange, and the electronics box.

The radiometric system was calibrated against the amplitude of the Gaussian fits to the Moon-signal profiles shown in Fig.~\ref{Fig3ini}, which we compared 
with the expression of Krotikov \& Pelyushenko (\cite{KroPel87}) for the brightness temperature at the center of the lunar disk at 2.3\thinspace{GHz}\ according 
to 
\begin{equation}\label{Tmoon}
T_\mathrm{MB}^\mathrm{Moon} = 221.63+3.14\cos\left(\Omega t -45\degr\right),
\end{equation}
where $\Omega t$ denotes the dependance on the optical phase of the Moon and sets the full Moon at $\Omega t = 0$. This expression is equivalent
to the one obtained by Mangum (\cite{Mangum93}), except that the one above allows for the averaging action of the antenna beam pattern. A
summary of calibration constants and antenna parameters obtained with this method is included in Table \ref{Tab2}.

\begin{figure*}
\centering\includegraphics[width=15cm]{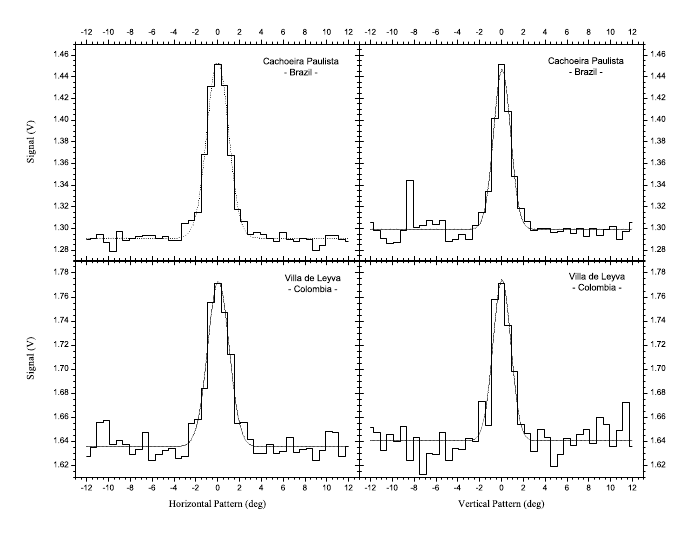}
\caption{Beam pattern characterization at 2.3\thinspace{GHz}\ from Moon measurements during routine survey scans away from the Galactic Plane.}
\label{Fig3ini}
\end{figure*}

There are some disadvantages with this method that impact the accuracy of our estimates, but we have assessed the uncertainties in several ways
to guarantee a reliable temperature scale for our map. Independent estimates of the total and main beam solid angles are the main source of
uncertainty, since the ratio of the beam efficiency, $\epsilon_M$, and the aperture distribution factor, $\kappa_p$, which scales the product of the
horizontal and vertical beam widths into the main beam solid angle, enter our gain calculation. Assuming a ratio of
$\epsilon_M/\kappa_p=0.75/1.05$ we have tested the validity of the Moon calibration method by factoring out the total beam solid angle from its
relation to the effective aperture, such that
\begin{eqnarray}\label{flux}
S_\mathrm{Jy} &=& \pi R_\mathrm{Moon}^2{2k\over\lambda^2}
\left({T_\mathrm{MB} \over S_\mathrm{volts}^\mathrm{peak}-S_\mathrm{volts}^\mathrm{ref}}\right)_\mathrm{Moon}\nonumber\\
&&\times\left({S_\mathrm{volts}^\mathrm{peak}\over 1 - \eta_\mathrm{G}\Delta T_\mathrm{cal}^\mathrm{peak}}
-{S_\mathrm{volts}^\mathrm{ref}\over 1 - \eta_\mathrm{G}\Delta T_\mathrm{cal}^\mathrm{ref}}\right)_\mathrm{source}\times 10^{26},
\end{eqnarray}
where $S_\mathrm{Jy}$ is the flux density of an observed known source in janskies, $R_\mathrm{Moon}$ is the angular radius of the lunar disk in radians,
$S_\mathrm{volts}$ is the signal output of the radiometer in volts, $\eta_\mathrm{G}$ is the thermal gain susceptability in Table \ref{Tab2}, and
$\Delta T_\mathrm{cal}$ is the temperature drift of the RF plate from its mean temperature during the Moon
$S_\mathrm{volts}^\mathrm{peak}$ measurements. We chose Vir A and Cen A to test the calibrating Moon amplitude in the Colombian and
Brazilian data sets, respectively. Vir A is sufficiently smaller than our beam size and we obtained 134\thinspace{Jy} using Eq.~\ref{flux} compared to
140\thinspace{Jy} using the Baars et al.~(\cite{Baars77}) semi-absolute spectrum between 400\thinspace{MHz} and 25\thinspace{GHz} as given by
\begin{eqnarray}\label{VirA}
\log S_\mathrm{Vir A} [\mathrm{Jy}] = 5.023(\pm0.034) - 0.856(\pm0.010)\log\nu_\mathrm{MHz}.
\end{eqnarray}
The Baars scale was updated by Ott et al.~(\cite{Ott94}) to assess source variability, and they obtained new measurements at $\lambda=21$, 11, 6,
and 2.8\thinspace{cm} for Vir A, which allowed them to fit a second-order polynomial between 1408\thinspace{MHz} and 10.55\thinspace{GHz}, such that
\begin{eqnarray}\label{VirAA}
\log S_\mathrm{Vir A}^\mathrm{Ott} [\mathrm{Jy}] = 4.484 - 0.603\log\nu_\mathrm{MHz} - 0.028\log^2\nu_\mathrm{MHz}.
\end{eqnarray}
However, at our frequency of 2.3\thinspace{GHz}\ this revised scale produces 138\thinspace{Jy}, which does not differ significantly from the Baars scale.
Cen A, on the other hand, is an extended source covering some $4\degr\times10\degr$ in celestial coordinates, so we targeted its brightest component,
the northeast inner lobe, and obtained 94\thinspace{Jy}, compared to 99\thinspace{Jy} using the Alvarez et al.~(\cite{Alvarez00}) integrated flux density relation
between 80\thinspace{MHz} and 43\thinspace{GHz} as given by
\begin{eqnarray}\label{CenA}
\log S_\mathrm{Cen A} [\mathrm{Jy}] = 4.35(\pm0.08) - 0.70(\pm0.02)\log\nu_\mathrm{MHz}.
\end{eqnarray}
The less than 5\% discrepancy between our estimates of both sources and the published values lends substantial support to the use of the observed
amplitude of the Moon signal to calibrate our antenna temperature scale. The ratio $\epsilon_M/\kappa_p$ does not enter
Eq.~\ref{flux} explicitly, but it is implied in the calculation of the RF plate physical temperature drift. Given the good agreement between observed and
published fluxes, the assumed ratio restrains the beam efficiency to the range defined by the uncertainty quoted in Table \ref{Tab2} on behalf of
the expected range for $\kappa_p$ alone ($1.00\le\kappa_p\le1.05$).

An additional source of uncertainty with the Moon calibration method is the need to estimate the temperature of the reference background level
above which the Moon signal is superimposed. This level consists of two isotropic components, the CMB radiation and a small contribution of the
diffuse background of extragalactic emission, but it also includes Galactic foreground radiation, atmospheric emission, and stray radiation from the
ground through the sidelobes. The CMB contribution has been measured accurately to be $2.725\pm0.002$\thinspace{K} (Mather et al.~\cite{Mather99}),
while we estimated the extragalactic background at $0.027$\thinspace{K} (Lawson et al.~\cite{Lawson87}). Likewise, we estimated the atmospheric emission at
$1.64\pm0.07$\thinspace{K} for the Colombian site and at $2.29\pm0.13$\thinspace{K} for the Brazilian site according to the modeling prescription of Danese \&
Partridge (\cite{DanPar89}). As for the Galactic foreground component, the observations of the Moon took place in regions away from the Galactic
Plane, so that an estimate of $0.8\pm0.4$\thinspace{K} was used in agreement with successive refinements of the calibration constants during the removal
of the stray radiation component. The assessment of ground contamination proved to be the single most challenging aspect in the preparation of the
GEM 2.3\thinspace{GHz}\ map, as will be seen in the next section, and for this reason the system temperature estimate given in Table \ref{Tab2} includes the
contribution of the ground.

\begin{table*}
\centering
\caption[]{GEM 2.3\thinspace{GHz}\ survey}
\begin{tabular}{lcc}
\noalign{\vskip -6pt}
\hline\hline
\noalign{\vskip 2.5pt}
\multicolumn{1}{c}{Description} & \multicolumn{1}{c}{Colombia} & \multicolumn{1}{c}{Brazil}\\
\noalign{\vskip 2pt}	 
\hline	 
\noalign{\vskip 2.5pt}	 
Observational site & Villa de Leyva & Cachoeira Paulista\\
Longitude (WGS84) & $-73\degr\,35\arcmin\,0.53\arcsec$ & $-44\degr\,59\arcmin\,54.34\arcmin$\\
Latitude (WGS84) & $+5\degr\,37\arcmin\,7.84\arcsec$ & $-22\degr\,41\arcmin\,0.74\arcsec$\\
Altitude (m.a.s.l.) & 2173 & 572\\
Observing runs & 1995-Jun-1--18 & 1999-May-18 -- Jun-17\\
 & & 1999-Oct-11--26\\
Antenna mounting  & altazimuthal  & altazimuthal\\
Azimuth scanning speed (rpm)& $0.99632\pm0.00036$ & $1.00290\pm0.00063$\\
Sky coverage (\%) & 46.3 & 46.8 \\
Pointing accuracy & $6\farcm84$ & $5\farcm26$ \\ 
Center frequency (MHz) & 2300  & 2300\\
Pre-detection BW (MHz) & 100  & 100 \\
Gain (\mbox{K V$^{-1}$})   & 54.675 & 50.928\\
Gain susceptability (K$^{-1}$)  & $-0.02381\pm0.00028$  & $-0.02164\pm0.00006$ \\
System temperature (K) & 85.466 & 61.644\\
RMS sensitivity (mK) & 11.42  & 8.24\\
RF plate $T_\mathrm{cal}$ (K)  & 310.572 & 308.031\\
Horizontal HPBW ($\degr$) & $2.30\pm0.13$  & $2.31\pm0.03$\\
Vertical HPBW ($\degr$) & $1.92\pm0.18$  & $1.82\pm0.12$\\
Beam efficiency (\%)  & $75.0\pm3.5$  & $75.0\pm3.5$\\
Aperture efficiency (\%) & 38.0  & 39.9\\
PSS \mbox{(Jy K$^{-1}$)} & 306 & 291\\
\noalign{\vskip 2.5pt}
\hline
\noalign{\vskip 2.5pt}
\end{tabular}
\label{Tab2}
\end{table*}

\section{Data description and processing}\label{DDaP}
The data used in the preparation of the GEM 2.3\thinspace{GHz}\ map were collected during three observational seasons at the Colombian and Brazilian sites 
listed in Table \ref{Tab1}. A brief description of each observational site and their associated data sets is given in the first part of Table \ref{Tab2}. 
We used $Z=30\degr$ scans for a total combined sky coverage of $\approx66$\%. In Fig.~\ref{Fig5} we have summarized the major steps in the preparation
of the combined map.

\subsection{Time stamping and pointing calibration}
The Colombian data set had each frame tagged in seconds and we adjusted a simple linear progression with a step size of $\tau$ seconds to
time-stamp each frame. For the Brazilian data, we relied on an iterative process to minimize the difference between the timing of the frames in the
timeline of 1\thinspace{s} updates of the GPS receiver and the expected cadence of the seamless data stream. We broke the latter into equal time 
slots of $2^\mathrm{h}$ and $34^\mathrm{m}$ to facilitate detection and correction of systematics during the data collection. Shorter time slots 
were inevitable whenever interruptions of the scanning process were encountered in lining up the TOD or when the data stream was found to be corrupted.

Readings from the azimuth angle encoder were first scaled to match rotations of 360\degr\ at constant speed. The ephemeris of the Sun was
used to estimate the offset between the astronomical azimuth of the Sun and the encoder readout of the radiometer peak signal during scans
with closest approach in the direction of the Sun. To guarantee a consistent mechanical configuration for all scans, we used a
calibrated steel bar to lock the dish into pre-defined elevation angles. Fine-tuning of the actual azimuth and elevation angles was accomplished by
re-projecting the polar coordinates, $\theta$ and $\phi$, of the boresight onto a celestial grid centered on the ephemeris-specified position of the 
Moon, as shown in Fig.~\ref{Fig6}. We estimated the pointing accuracy of the surveys obtained from the Colombian and Brazilian data using the RMS 
deviation of the centroid of the Gaussian fits in Fig.~\ref{Fig3ini} from the origin of the Moon-centered coordinate system. We obtained RMS values
of $6\farcm84$ from the Colombian data and $5\farcm26$ from the Brazilian data.

\subsection{Total power systematics}
Once the time-ordered data were properly synchronized, we extracted the raw signal of the Moon and calibrated the radiometer
constants as described in Sect.~\ref{RCH}. Of these constants, the thermal gain susceptability $\eta_\mathrm{G}$ required extraction of the raw
signal observed in pre-defined regions of low Galactic emission, or cold sky regions, as shown in the 408\thinspace{MHz}\ survey of Haslam et al.~(\cite{Haslam82}), 
to monitor the linear response of the radiometer signal output as a function of the RF plate physical temperature. Fortunately, these low-emission 
regions were scanned during routine day-time hours, when the drift in $\Delta T_\mathrm{cal}$ displayed its largest excursions and, thus,
enabled accurate linear fits to the signal-RF plate temperature correlation. A set of three regions of $5\degr\times 5\degr$ were chosen for each site as
listed in Table \ref{Tab4} and the linear correlation coefficients of each set were averaged and scaled by the corresponding ratio of gain to system
temperature to obtain $\eta_\mathrm{G}$.

\begin{table}
\caption[]{Low Galactic emission regions}
\begin{tabular}{ccc}
\noalign{\vskip -6pt}
\hline\hline
\noalign{\vskip 2.5pt}
Region&\multicolumn{2}{c}{Celestial coordinates -- Epoch 2000}\\
& \multicolumn{2}{c}{(degrees)}\\
\noalign{\vskip 2pt}
\hline
\noalign{\vskip 2.5pt}
Colombia 1 & $141.0\le\mathrm{RA}\le 146.0$ & $30.0\le\mathrm{DEC}\le 35.0$ \\
Colombia 2 & $134.5\le\mathrm{RA}\le 139.5$ & $28.5\le\mathrm{DEC}\le 33.5$ \\
Colombia 3 & $136.0\le\mathrm{RA}\le 141.0$ & $1.5\le\mathrm{DEC}\le 6.5$ \\
Brazil 1 & $45.0\le\mathrm{RA}\le 50.0$ & $-33.0\le\mathrm{DEC}\le -28.0$ \\
Brazil 2 & $54.0\le\mathrm{RA}\le 59.0$ & $-33.0\le\mathrm{DEC}\le -28.0$ \\
Brazil 3 & $37.0\le\mathrm{RA}\le 42.0$ & $-30.0\le\mathrm{DEC}\le -25.0$ \\
\noalign{\vskip 2.5pt}
\hline
\noalign{\vskip 2.5pt}
\end{tabular}
\label{Tab4}
\end{table}

Total power systematics are both linear and non-linear in nature. Thermal gain susceptability corrections address some of the systematics of the
linear type. Non-linear gain variations, on the other hand, cannot be corrected for without additional time-series analysis of the data stream. Initially,
we had intended to use the periodical firings of the noise source to correct gain variations in general. However,
after some careful analysis of the noise source amplitude variations, we found that the noise source had become unstable in the course of time and
that its use as a transfer calibrator failed to meet the accuracy we required. To filter oddly behaving data, we turned to an alternate criterion,
which relied on the distribution of the lowest observed antenna temperature $T_\mathrm{A}^\mathrm{min}$ per time slot
($154^\mathrm{m}$ max). Non-linear excursions of the baseline were found to correspond to low values of $T_\mathrm{A}^\mathrm{min}$, so we
fixed lower limits for its distribution to improve the characterization of the systematics that affect the survey.

\begin{figure*}
\centering\includegraphics[width=15cm]{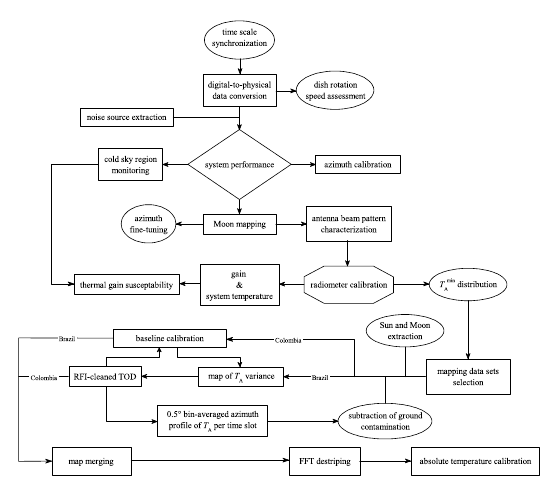}
\caption{Overview of the survey data processing.}
\label{Fig5}
\end{figure*}

For this purpose, we split the Brazilian survey into three separate data sets, each comprising about two weeks of data collection, whose characteristics
reflect the seasonality effect between summer and winter, but also the long-term performance of the experiment over a month of winter observations.
The first of the two Brazilian winter data sets (winter I) coincided with the same period of the year as the single Colombian data set. Fig.~\ref{Fig7} 
shows the distribution of $T_\mathrm{A}^\mathrm{min}$ for the four data sets. All three Brazilian data sets are seen to be well contained under a single-peaked
distribution centered around 4.5\thinspace{K}, whereas the Colombia data are not as smoothly distributed with a main peak below 3\thinspace{K}, a secondary
peak coinciding with the single peak of the Brazilian data sets, and a likely third and smaller peak around 6\thinspace{K}. Although a realistic lower 
limit at 3.5\thinspace{K} could be chosen for the Brazilian data, as indicated by the vertical arrow in the figure, a value of 2\thinspace{K} had to be used 
for the Colombian data set to avoid excessive decimation of the survey.

The clumpy distribution of $T_\mathrm{A}^\mathrm{min}$ in the Colombian data set does have an immediate effect in the mapping of its survey data. 
As Fig.~\ref{Fig8} shows, there is clear evidence of an inhomogeneous baseline due to scan-induced circular striping. On the other hand, mapping of
the Brazilian survey shows a smoother baseline contrast, but a rather different and disturbing systematic shows up in the form of heavy striping in
declination as a result of substantial ground pick-up in the northern part of the Cachoeira Paulista sky.

\subsection{RFI and ground contamination removal}
The two maps in Fig.~\ref{Fig8} show several bright spots due to man-made interference (RFI), whose excision was accomplished by means of a
low-pass filter during a four-step iteration process aimed at removing the ground striping. In the Colombia map there is also the sinusoidal streak left by
an artificial satellite, whose transmissions highlight portions of the contaminating track. To remove this type of RFI, a tedious monitoring of the
survey time slots was conducted and the affected time slots were cleaned of observations found within the perimeter of the satellite streak.

The cleaning of the ground striping proved, on the other hand, to be a homogeneous baseline-dependant process; so it was applied iteratively to the
Colombian and Brazilian data sets in slightly different ways as indicated at the bottom of Fig.~\ref{Fig5}. For both data sets, the process started by
flagging the presence of the Sun and Moon whenever their angular separation from the observing direction was less than $30\degr\to 60\degr$ and
$6\degr$, respectively. Subsequently, a variance map of the antenna temperatures in the remainder of the survey was obtained; but the
inhomogeneous baseline of the Colombian data set required its baseline to be calibrated as described in Sect.~\ref{MAP} before its variance map
was obtained. In a second step, the observations were compared to the variance map on a per pixel basis and a new set of TOD was assembled from
those observations that did not exceed three times the variance in the corresponding pixel. A pixel resolution of $360\degr/256 = 1\fdg40625$ was
chosen to prevent undersampling for the size of our data sets. In the third iteration step, the new and RFI-cleaned TOD were averaged in
azimuth to produce mean profiles of the ground contamination for each observed time slot, as long as the observations were distanced more than
$30\degr$ from the Galactic Plane to prevent real sky features from appearing as contaminating signatures in the mean azimuth profile. Finally, in the last
and fourth step, an overall mean azimuth profile for each of the four data sets was secured by excluding any remaining isolated sky feature. This was
accomplished with a low-pass filter, which looped over the azimuth profiles from all time slots in a given set, to exclude profile points with more
than three times the variance per azimuth bin ($0.5\degr$ wide). The loop was repeated until no more profile points were excluded. From here on, the
four-step iteration process was repeated, but the subtraction of the normalized overall-mean azimuth profile from the previous iteration was
subtracted in the first two steps. The normalization 
\begin{figure*}
\centering\includegraphics[width=15cm]{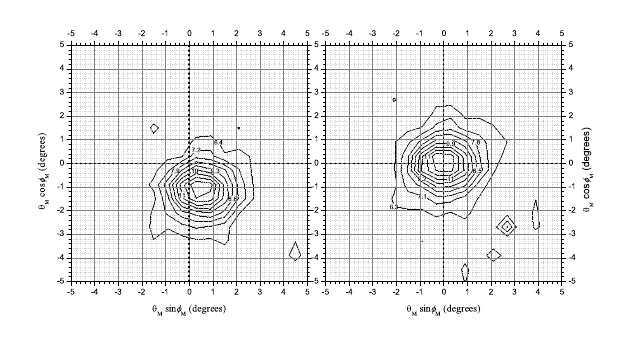}
\caption{Images of the Moon before (left panel) and after (right panel) fine-tuning the pointing calibration constants. Contour levels are 
in antenna temperature units.} 
\label{Fig6}
\end{figure*}
consisted in offsetting the profile down by the amplitude of its lowest profile point. In this way,
the ground contamination was successively eliminated until the overall-mean aziumth profile of the last iteration was practically reduced to zero.

\begin{figure}
\resizebox{8.8cm}{!}{\includegraphics{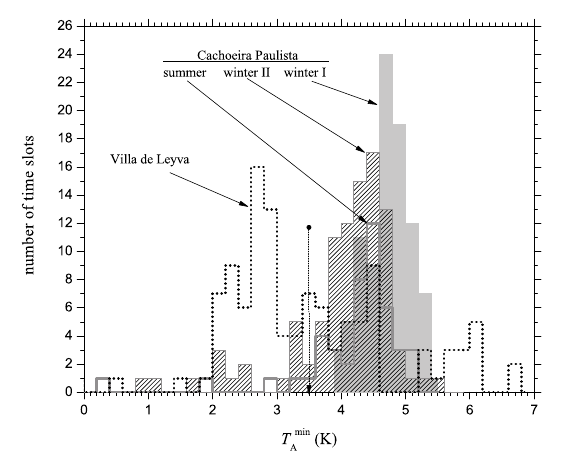}}
\caption{Histograms of observed antenna temperature minima in the Colombian and Brazilian data sets: Villa de Leyva (dotted line) and Cachoeira 
Paulista (gray line -- summer; gray-filled -- winter I; hatched -- winter II).}
\label{Fig7}
\end{figure}

\begin{table}
\caption[]{Surveyed data composition.}
\begin{tabular}{lrrcrr}
\noalign{\vskip -6pt}
\hline\hline
\noalign{\vskip 2.5pt}
& \multicolumn{2}{c}{Colombia} && \multicolumn{2}{c}{Brazil} \\
\noalign{\vskip 2pt}
\cline{2-3}\cline{5-6}
\noalign{\vskip 2.5pt}
\multicolumn{1}{l}{Description} & \multicolumn{1}{c}{(hours)} & \multicolumn{1}{c}{(\%)}&
& \multicolumn{1}{c}{(hours)} & \multicolumn{1}{c}{(\%)} \\
\noalign{\vskip 2pt}
\hline
\noalign{\vskip 2.5pt}
Survey total & 162.85 & 100.00 && $532.11$ & 100.00 \\
Corrupted data & 4.13 & 2.54 && ... & ... \\
TOD mismatch & 6.56 & 4.03 && ... & ... \\
Noise source & 11.41 & 7.01 && 39.91 & 7.5 \\
Sun and Moon & 17.0 & 10.44 && 78.11 & 14.68 \\
$T_\mathrm{A}^\mathrm{min}$ exclusion & 4.38 & 2.69 && 43.36 & 8.15 \\
RFI & 2.46 & 1.51 && 3.46 & 0.65 \\
\noalign{\vskip 2.5pt}
\hline
\noalign{\vskip 2.5pt}
Mapping total & 116.91 & 71.79 && 367.27 & 69.02 \\
\noalign{\vskip 2.5pt}
\hline
\noalign{\vskip 2.5pt}
\end{tabular}
\label{Tab5}
\end{table}

Fig.~\ref{Fig9} shows the cumulative overall-mean azimuth profiles of the four data sets after meeting the convergence criterion. The spiky
Brazilian profiles reflect the nature of the striping in declination, but they also highlight the seasonality effect and the need to assess winter and
summer data sets separately. As expected, the profile of the Colombian data set shows the contamination from the ground to be much lower and less
variable. Still, residual horizontal striping remained visible in the final iterated map and additional destriping was applied in the final preparation of
the map as described in the next section.

Table \ref{Tab5} summarizes all cuts applied to the data sets as a result of the selection and cleaning processes described so far. A more
conservative minimum angular separation of $60\degr$ for the Sun was used with the Brazilian data sets to eliminate traces of scattering sidelobes from
the three-legged feed support structure as well as to guarantee a safer margin of secondary sidelobe suppression. The angular separation for the
Colombian data set at $30\degr$ reflects the lack of conclusive evidence for sidelobe contamination at larger angular separations due to the reduced
sensitivity from a smaller data volume. Corrupted data and TOD mismatch did not apply to the Brazilian data sets, since an automatic data collection
algorithm was implemented to keep the observational time slots of equal length. 
\begin{figure*}
\centering\includegraphics[width=18cm]{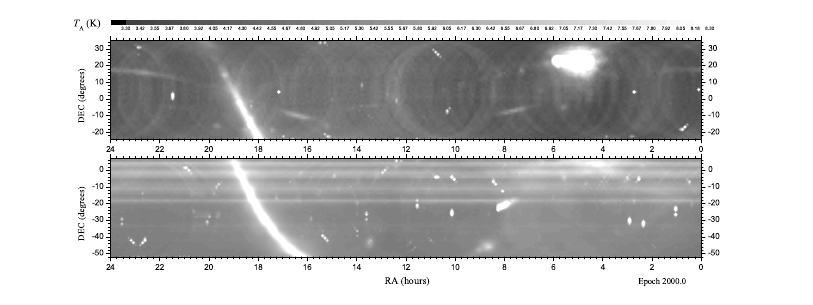}
\caption{Maps of unprocessed survey data from selected time slots ($T_\mathrm{A}^\mathrm{min}$ limited) in the Villa de Leyva (top panel) and 
in Brazil's winter I (lower panel) data sets, including the prominent feature introduced by the Sun around $5^\mathrm{h}$ of RA in the Colombian map.}
\label{Fig8}
\end{figure*}
\begin{figure}
\resizebox{8.8cm}{!}{\includegraphics{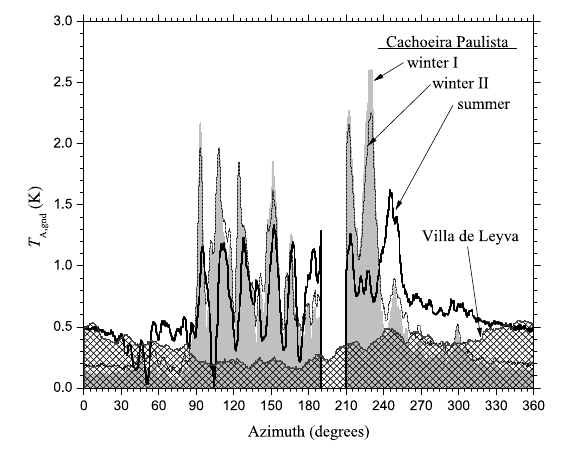}}
\caption{Cumulative overall-mean azimuth profiles of the ground contamination component of the antenna temperature: Villa de Leyva (cross-hatched)) 
and Cachoeira Paulista (thick line -- summer; gray-filled -- winter I; thin line -- winter II).}
\label{Fig9}
\end{figure}
Finally, some 68 hours of data associated with extreme weather conditions, severe cases of RFI and unusually large excursions in the readings of the temperature sensors and of the radiometer signal were flagged for exclusion during the observing runs in Colombia and did not enter the processing scheme outlined in Fig.~\ref{Fig5}. Consequently, they were not accounted for as part of the total survey time in Table \ref{Tab5} when compared with Table \ref{Tab1}.

\section{Final map preparation and calibration}\label{MAP}
The final preparation of the GEM 2.3\thinspace{GHz}\ survey was divided into three major steps. First, the baseline of the ground-subtracted Brazilian data 
sets was calibrated and the resulting map was merged with the one from the cleaned Colombian data set. Second, the destriping technique developed by
Platania et al.~(\cite{Platania03}) was applied to the merged map to clean residual striping due to ground contamination and baseline inhomogeneities. Third, 
a direct comparison with the Rhodes/HartRAO survey at 2326\thinspace{MHz} (Jonas et al.~\cite{Jonas98}) was performed to absolutely calibrate the GEM survey.

\subsection{Baseline calibration}
To reduce the circular stripes introduced by the scanning technique in the TOD of the Brazilian data set, its baseline was calibrated following
the same prescription as was applied to the Colombian data set (see Fig.~\ref{Fig5}). This calibration consisted in resolving the true sky
temperature distribution at the level where its coldest contour would intersect, at least once, every survey scan (a $60\degr$ wide circular scan
centered at the zenith with a $2\fdg30\times1\fdg85$ HPBW). In other words, the coldest observation of every survey scan would be normalized to a
uniform temperature, or apparent zero-point of our temperature scale, across the entire declination band. This assumption is, to a first-order
approximation, sufficient to correct the 1/f-noise in the TOD as the dominant source of baseline inhomogeneity, but it also defines an effective
full-beam brightness temperature scale for the survey.

When applied to the Colombian data set, the precision of the baseline calibration was set by the level of the fluctuations in the ground profile of
Fig.~\ref{Fig9}, or $62\pm16$\thinspace{mK}. Without this {\it a priori} calibration, the ground profile fluctuations would have increased to
$1357\pm424$\thinspace{mK} and would have compromised any subsequent baseline calibration. For comparison, the ground profile
fluctuations in the Brazilian data averaged to $355\pm55$\thinspace{mK} in the winter I, $350\pm56$\thinspace{mK} in the winter II and $362\pm57$\thinspace{mK} in the summer data sets before their baseline was calibrated. This difference in implementing the baseline calibration did not allow the Colombian and
Brazilian surveys to be merged along their entire region of overlap ($-23\degr\la\delta\la+6\degr$). Instead, an offset of $148\pm24$\thinspace{mK} was applied
to the Colombian data according to the difference between the two surveys along the northern boundary (best signal-to-noise ratio) of the Brazilian
survey for $\alpha\la17$\thinspace{h} and around an overlapping rectangular region $\sim11\degr$ wide for $\alpha\ga17$\thinspace{h}. The merging of the two
surveys in this overlap region allowed resiliant traces of ground striping in the Brazilian survey to be reduced. Ultimately, the baseline calibration
relied on the uniformity in the sampling of the surveyed sky area. With our observational technique of zenith-centered circular scans, the sky is
sampled preferentially toward the edges of the declination band. 
\begin{figure*}
\centering\includegraphics[width=18cm]{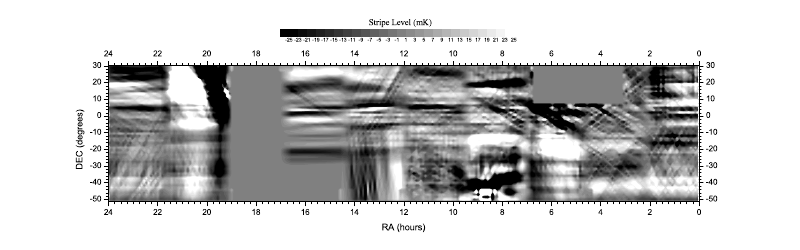}
\caption{Map of residual striping obtained with the FFT filtering technique of Platania et al.~(\cite{Platania03}) on the ground-corrected merged map.}
\label{Fig10}
\end{figure*}
This intrinsic bias diminishes with the number of observations by statistical virtue
alone and, thus, contributed to the baseline differences between the two maps. Of course, cuts in the TOD, such as those indicated in
Table \ref{Tab5}, also affected the sampling uniformity and introduced additional baseline inhomogeneities.

\subsection{Cleaning of residual striping}\label{residual}
The merged map showed traces of residual striping, whose RMS level could be estimated at 58\thinspace{mK} after filtering the zero-frequency component
of the map in Fourier space. The FFT destriping technique of Platania et al.~(\cite{Platania03}) was subsequently applied to produce a cleaner map.
The technique was adapted from a method applied by Schlegel et al.~({\cite{Schlegel98}) to destripe the IRAS map and uses source extraction and
stripe identification thresholds in 30 partially overlapping patches of $32\times 32$ pixels for the GEM map. This small number of pixels per patch
and the corresponding large angular dimension was suitable to address the variability of the striping pattern in the merged map and resulted in the
stripe image shown in Fig.~\ref{Fig10} with an RMS level of 22\thinspace{mK}. The procedure was not applied to five patches (15.1\% of the map) containing
the Galactic Plane because of steep signal gradients and a clean GEM map was obtained by subtracting the stripe image from the merged map.

The destriping procedure was applied in Platania et al.~(\cite{Platania03}) to the 408\thinspace{MHz}\ full-sky map of Haslam et al.~(\cite{Haslam82}) and to 
the 1420\thinspace{MHz} (Reich \cite{Reich82}, Reich \& Reich \cite{Reich86})) and 2326\thinspace{MHz} (Jonas et al.~\cite{Jonas98}) surveys. Given the
importance of the latter for the absolute calibration of the GEM map, we compare in Table \ref{Tab6} the effect of the destriping on the GEM and
Rhodes maps in terms of the percentage temperature variations $\delta T$ of the pixels. Nearly half of the destriped pixels in the GEM map did
not exceed variations of more than 5\%, whereas in the Rhodes map the corresponding percentage of pixels was higher ($\approx70$\%). Therefore, we 
examined the significance of the variations larger than 5\% for the GEM map, and Table \ref{Tab6} also shows the distribution of the percentage
variations in two separate temperature regimes, GEM-I of negative residuals ($\Delta T < 0$\thinspace{K}) and GEM-II of positive residuals 
($\Delta T > 0$\thinspace{K}).
Despite the higher contribution of pixels with positive residuals with variations $\delta T > 10$\%, their average residual was only $27\pm17$\thinspace{mK},
whereas the smaller fraction of pixels with negative residuals in the same range averaged to a significantly higher negative residual of $-51\pm44$\thinspace{mK}. For variations in the 5\%$< \delta T \le 10$\% range, the average residuals were $13\pm11$\thinspace{mK} and $-18\pm13$\thinspace{mK}. 
Altogether, this tendency of the negative residuals to require larger corrections than the positive residuals could imply that the ground subtraction slightly
overcorrected this type of contamination.

\begin{table}
\caption[]{Distribution of percentage temperature variations $\delta T$ in destriped pixels of the GEM and Rhodes/HartRAO surveys.}
\begin{tabular}{crrrr}
\noalign{\vskip -6pt}
\hline\hline
\noalign{\vskip 2.5pt}
\noalign{\vskip 2.5pt}
Survey & \multicolumn{1}{c}{$<1$} & \multicolumn{1}{c}{$1 < \delta T < 5$} & \multicolumn{1}{c}{$5 < \delta T < 10$} & \multicolumn{1}{c}{$>10$} \\
& (\%) & \multicolumn{1}{c}{(\%)} & \multicolumn{1}{c}{(\%)} & (\%) \\
\noalign{\vskip 2pt}
\hline
\noalign{\vskip 2.5pt}
GEM-I & 6.5 & \multicolumn{1}{c}{22.1} & \multicolumn{1}{c}{13.9} & 6.0 \\
GEM-II & 5.9 & \multicolumn{1}{c}{17.2} & \multicolumn{1}{c}{12.0} & 16.4 \\
GEM & 12.4 & \multicolumn{1}{c}{39.3} & \multicolumn{1}{c}{25.9} & 22.4 \\
Rhodes & 18.3 & \multicolumn{1}{c}{50.6} & \multicolumn{1}{c}{22.5} & 8.6 \\
\noalign{\vskip 2.5pt}
\hline
\noalign{\vskip 2.5pt}
\end{tabular}
\label{Tab6}
\end{table}

\subsection{Absolute calibration}
The absolute baseline calibration was obtained by means of a {\it T-T} plot between destriped versions of the GEM and Rhodes maps. The
Rhodes survey, from which the isotropic components of the CMB and the diffuse extragalactic background had been subtracted, was convolved with a
Gaussian profile with a FWHM of $140\arcmin$ to match the GEM survey. The overlap between the two surveys covers the declination range
$-51\degr\la\delta\la+30\degr$ after excluding boundary systematics. Using a celestial grid with a resolution of $360\degr/256 = 1\fdg40625$ produced
13\,132 paired pixels, whose {\it T-T} plot is shown in Fig.~\ref{Fig11} along with a linear fit (98\% correlation) for calibrating the true zero-point
of the GEM survey baseline. 

Notwithstanding the spectral implications due to the small deviation of the linear coefficient from unity
($+14.3\pm1.6\,\mathrm{mK~K}^{-1}$ is equivalent to a temperature ratio between two synchrotron components, had the GEM band been centered
at $2314.05\pm0.45$\thinspace{MHz}), we can alternatively estimate the zero-point correction from the straight mean of the pixel-by-pixel temperature
differences at $3.125\pm0.092$\thinspace{K}. Similarly, the distribution of these pixel differences is nearly normal, as shown in Fig.~\ref{Fig12}, with a
Gaussian envelope centered at $3.1356\pm0.0019$\thinspace{K} and standard deviation of $65.5\pm1.9$\thinspace{mK}. An important feature of the destriping
procedure is that it has only a negligible effect on the temperature scale of the map, and to illustrate this, we included in Fig.~\ref{Fig12} the Gaussian
envelope and its underlying distribution of pixel differences for the ground-corrected map before the destriping. We estimate the error in the
zero-level of our survey at 103\thinspace{mK} by combining in quadrature the standard deviation of the pixel differences with the error of  80\thinspace{mK} 
in the absolute zero-level of the Rhodes survey, according to their calibration with the absolute sky temperature measurements of Bersanelli 
et al.~(\cite{Bersanelli94}) at 2\thinspace{GHz} from the South Pole. 

\begin{figure}
\resizebox{8.8cm}{!}{\includegraphics{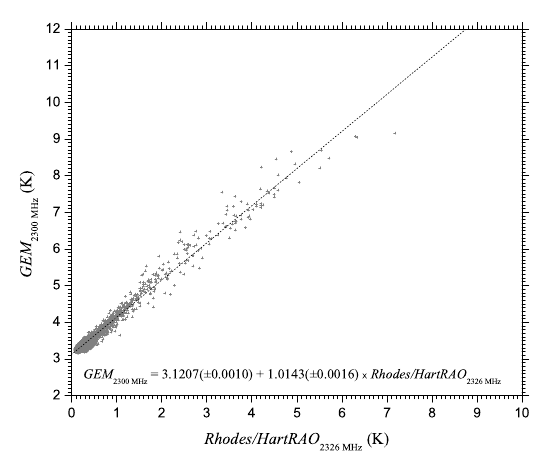}}
\caption{{\it T-T} plot between the destriped GEM survey at 2300\thinspace{MHz} and the Rhodes/HartRAO survey at 2326\thinspace{MHz}.}
\label{Fig11}
\end{figure}
\begin{figure}
\resizebox{8.8cm}{!}{\includegraphics{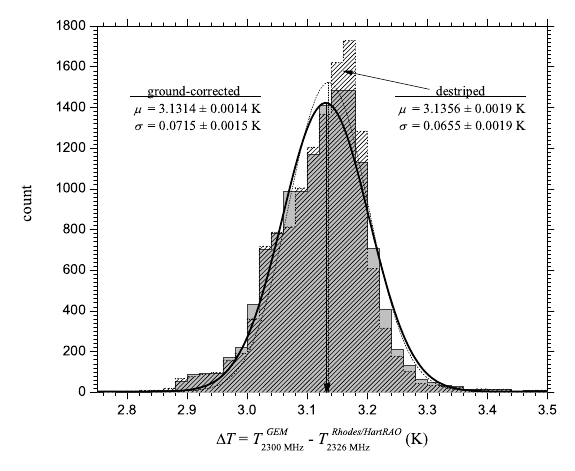}}
\caption{Distribution of pixel-by-pixel temperature differences between GEM and Rhodes maps before (ground-corrected) and after the destriping.}
\label{Fig12}
\end{figure}

Our estimate agrees with the expected accuracy of the Moon-calibrated
temperature scale in Sect.~\ref{RCH}, where we quoted an error $< 5$\% that the cold sky baseline calibration translates into $< 156$\thinspace{mK}.
Similarly, the difference between the apparent Moon calibration-dependant zero-point and its true value is 108\thinspace{mK}.

The clean and calibrated GEM survey at 2.3\thinspace{GHz}\ is presented in Fig.~\ref{Fig13}. The brightness temperature scale of the survey is tied up to  
the resolution-limited approach of the cold sky baseline calibration and the full-beam definition of the Rhodes survey. We can obtain a conversion factor 
between the full-beam estimates of the two surveys by linearly correlating the GEM survey with the unconvolved version of the Rhodes survey. This gives 
a conversion factor $0.8151\pm0.0031$, which matches the 0.8641 for the emission ratio in the direction of the Galactic Center reasonably well,
given the 5\% uncertainty of the Rhodes temperature scale and the compounded error in the zero-level accuracy of the GEM survey. In Figs.~\ref{Fig14} 
and~\ref{Fig15} a set of pre-defined contour levels shows a direct comparison between these two surveys, which enables some common features to be easily
identified. In the next section we address some differences that emerge in the morphology of their large-scale structure of the radio continuum. 

\begin{figure*}
\centering\includegraphics[width=18cm]{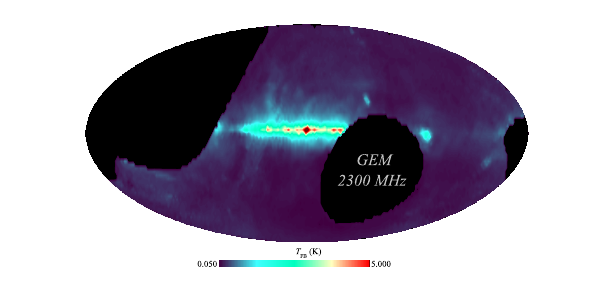}
\caption{Mollweide projection of the 2.3\thinspace{GHz}\ GEM survey in Galactic coordinates.}
\label{Fig13}
\end{figure*}

\begin{figure*}
\centering\includegraphics[width=10.5cm]{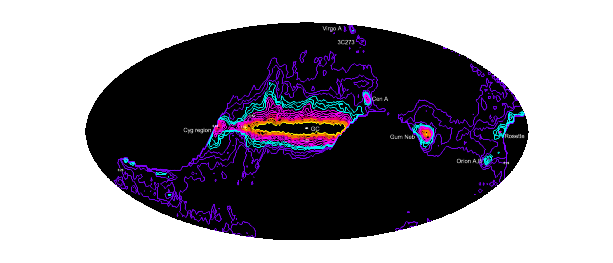}
\caption{Mollweide projection in Galactic coordinates of the large-scale structure of the radio continuum in the 2.3\thinspace{GHz}\ GEM survey according to selected contours at the level of detection of well-known
radio sources and Galactic features: (a) Virgo A and 3C273 contours between 0.23 and 0.43\thinspace{K}; (b) Orion A, B and Rosette Nebula contours between 0.43 and 
0.63\thinspace{K}; (c) Centaurus A and outer countours of the Cygnus region between 0.64 and 1.03\thinspace{K}; (d) inner contours of the Cygnus region between 
1.03 and 1.31\thinspace{K}; and (e) innermost contours of the Gum Nebula.}
\label{Fig14}
\end{figure*}

\begin{figure*}
\centering\includegraphics[width=10.5cm]{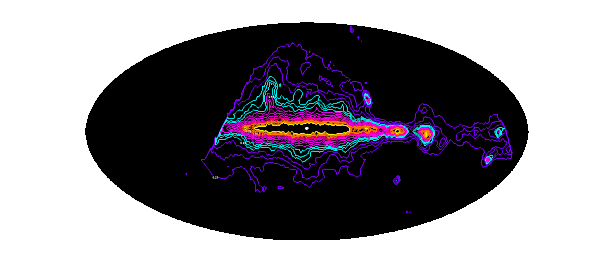}
\caption{Mollweide projection in Galactic coordinates of the large-scale structure of the radio continuum in the Rhodes/HartRAO survey at 2326\thinspace{MHz} according to the same selected contours as in Fig.~\ref{Fig14}.}
\label{Fig15}
\end{figure*}

\section{Discussion}\label{RES}

In Sect.~\ref{GEM} we introduced the GEM project as an experiment conceived to improve the mapping of the spectral and large-scale properties of Galactic
synchrotron emission. This improvement would be centered on overcoming two major undesired radioastronomical shortcomings that affect existing ground-based
surveys: the mutual consistency of their baselines and the observational bias of striping effects. In this section we discuss preliminary results based on
the GEM 2.3\thinspace{GHz} survey which, when compared to the already published HartRAO/Rhodes survey in the same frequency band, indicate a promising trend
toward achieving these goals. 

Our discussion will be developed in the framework of a concept introduced by Bennett et al.~(\cite{Bennett03}) for defining foreground masks 
in studies of the CMB with the WMAP data products. Their basic assumption was that the asymmetry in a histogram of the sky temperature distribution is 
due to Galactic foreground emission. In their Figure 1, they show how the sky temperature profile for the first-year results of the WMAP mission in the 
K-band may be decomposed into a symmetric distribution about the peak and a remainder. Even neglecting the effect of point sources, this remainder is 
recognized as a temperature excess caused by the intervening presence of our Galaxy. 

\subsection{All-sky surveys}

Since the K-band survey of the WMAP mission has a significant synchrotron component of Galactic emission, we included in the present discussion
the 408\thinspace{MHz} all-sky survey of Haslam et al.~(\cite{Haslam82}) with its notably dominant synchrotron character. We chose the 
one-degree smoothed version of the latest co-added total intensity map in the K-band (http://lambda.gsfc.nasa.gov) as well as a destriped version of the 
408\thinspace{MHz} map (Platania et al.~\cite{Platania03}). Their equal-area Mollweide projections in celestial coordinates are displayed in Fig.~\ref{Fig16}. 
This choice of coordinates enhances the declination-dependant nature of ground-based surveys and, to make the results comparable across all  
surveys, they were convolved to a resolution of $2\fdg3$ for consistency with the GEM survey. Fig.~\ref{Fig17} shows the histograms of their 
corresponding temperature distributions. Their profiles are characterized by a similar two-component separation, which we applied to the color rendering 
of the maps, such that their shades of gray pixels represent the transition between the symmetric distribution of low-temperature pixels in black and the 
excess foreground emission within the bounds of the temperature scale of the maps. Conversely, pixels found in the range between the lower limit of the 
remainder and the upper limit of the symmetric distribution are displayed in gray.  

\begin{figure*}
\centering\includegraphics[width=12cm]{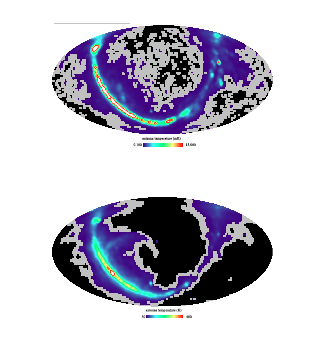}
\caption{Mollweide projection in celestial coordinates of the transition (gray pixels) between low- and high-emission regions in the all-sky surveys of the WMAP K-band at 23\thinspace{GHz} and of the Haslam map 
at 408\thinspace{MHz}.}
\label{Fig16}
\end{figure*}

\begin{figure*}
\centering\includegraphics[width=14cm]{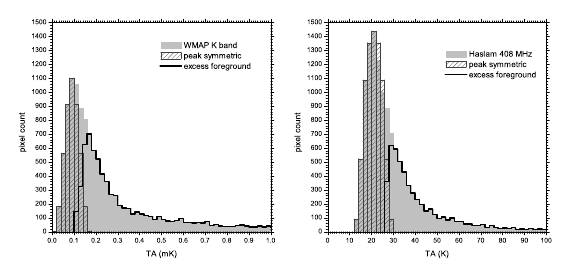}
\caption{Histograms of the sky temperature distributions in the maps displayed in Fig.~\ref{Fig16}.}
\label{Fig17}
\end{figure*}

The increasingly higher contrast of synchrotron radiation in Galactic emission toward the lower frequencies appears to be the determining factor of the 
appearance in the morphology of the transition region. At 408\thinspace{MHz}, the transition beyond the Galactic ridge marks a well-defined boundary between 
the high-latitude areas of low Galactic emission and the excess foreground emission from the Galactic Plane and Bulge, as well as from structures like the 
North Galactic Spur. At 23\thinspace{GHz}, the transition region in the WMAP K-band survey breaks up into several patches of gray pixels that blend without
any sharp boundary into the low-emission region of the symmetric component. From these considerations we accordingly expect that at intermediate frequencies 
the two-component scenario should be a constant feature of the large-scale structure of the radio continuum, and that the transition between high- and 
low-emission regions should disperse into the latter with increasing frequency.

\subsection{Partial sky coverage} 

The Rhodes and GEM surveys, given their partial-sky coverage, are not readily comparable with the all-sky surveys. Ground-based experiments with their 
sky-scanning limitations are confined to bands of declination centered on the zenith of the observational site. A quick inspection of the surveys in 
Fig.~\ref{Fig16} shows that any declination band will undoubtedly display the two-component separation, but their character may change with the range of 
Galactic Longitude that is intercepted. The Rhodes and GEM surveys were observed from observational sites with a privileged view of the central regions of 
the Galaxy. Consequently, we secured the common declination band in the range of  $-51\degr < \delta < +9\degr$ for comparison with the WMAP K-band 
and 408\thinspace{MHz} surveys. Below we refer to it as the GEM strip. 
\begin{figure*}
\centering\includegraphics[width=14cm]{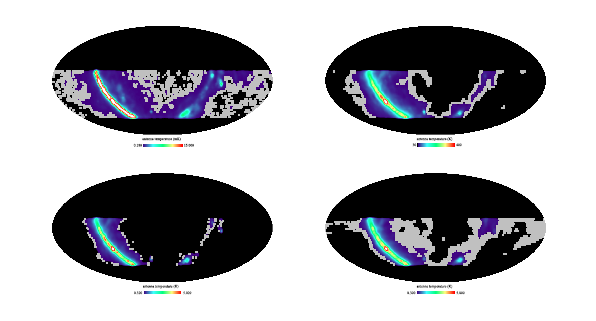}
o\caption{Mollweide projection in celestial coordinates of the transition regions for partial-sky coverage in the GEM strip: (upper left) WMAP K-band; (upper right) 408\thinspace{MHz}; (lower left) Rhodes; and (lower right) GEM.}
\label{Fig18}
\end{figure*} 

The upper panels in Fig.~\ref{Fig18} show the effect of restricting the sky coverage of the WMAP K-band and 408\thinspace{MHz} surveys to the GEM strip. 
Their histograms in the upper two panels of Fig.~\ref{Fig19} show the expected two-component distribution. Moreover, since the lower limit of the remainder 
and the upper limit of the symmetric component are precisely the same as those found for the all-sky versions, the gray-shaded pixels representing the 
transition between the two components in Fig.~\ref{Fig18} correspond to the exact same pixels in Fig.~\ref{Fig16}. We may therefore conclude that since the 
general properties of the morphology of the transition region in the WMAP K-band and 408\thinspace{MHz} surveys are not affected by the particular choice of 
sky coverage in the GEM strip, the Rhodes and GEM surveys can be tested for these same general properties.

\begin{figure*}
\centering\includegraphics[width=14cm]{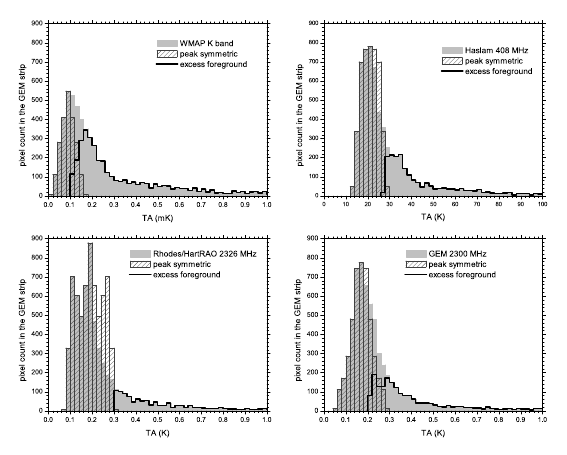}
\caption{Histograms of the sky temperature distributions in the partial sky maps displayed in Fig.~\ref{Fig18} showing their separation into symmetric and 
excess temperature components.}
\label{Fig19}
\end{figure*}

The results are shown in the lower panels of Fig.~\ref{Fig18} and \ref{Fig19}. Because the Rhodes and GEM surveys are mapping the sky essentially at the  
same frequency, the morphology of their transition zones should not be significantly different from each other. Yet, these results tell a different story. 

\begin{table*}[ht]
\caption[]{Spectral index in the 408\thinspace{MHz} and WMAP K-band emission regions.}
\centering
\begin{tabular}{cccccc}
\noalign{\vskip -6pt}
\hline\hline
\noalign{\vskip 2.5pt}
\multicolumn{2}{c}{Survey} && \multicolumn{3}{c}{Emission region in Survey 2} \\
\noalign{\vskip 2pt}
\cline{1-2}\cline{4-6}
\noalign{\vskip 2.5pt}
1 & 2 && high & transition & low \\
\noalign{\vskip 2pt}
\cline{1-2}\cline{4-6}
\noalign{\vskip 2.5pt}
WMAP K-band\tablefootmark{a} & 408\thinspace{MHz}\tablefootmark{a} && $-2.72\pm 0.19$ & $-2.85\pm 0.15$ & $-2.90\pm 0.14$ \\
\noalign{\vskip 2pt}
\cline{1-2}\cline{4-6}
\noalign{\vskip 2.5pt}
WMAP K-band\tablefootmark{b} &&& $-2.72\pm 0.19$ & $-2.81\pm 0.17$ & $-2.88\pm 0.14$ \\
Rhodes/HartRAO & 408\thinspace{MHz}\tablefootmark{b} && $-2.67\pm 0.11$ & $-2.66\pm 0.10$ & $-2.61\pm 0.10$ \\
GEM &&& $-2.72\pm 0.20$ & $-2.69\pm 0.18$ & $-2.56\pm 0.19$ \\
\noalign{\vskip 2pt}
\cline{1-2}\cline{4-6}
\noalign{\vskip 2.5pt}
408\thinspace{MHz}\tablefootmark{a} & WMAP K-band\tablefootmark{a} && $-2.72\pm 0.16$ & $-2.90\pm 0.07$ & $-3.02\pm 0.09$ \\
\noalign{\vskip 2pt}
\cline{1-2}\cline{4-6}
\noalign{\vskip 2.5pt}
408\thinspace{MHz}\tablefootmark{b} &&& $-2.72\pm 0.16$ & $-2.88\pm 0.17$ & $-3.01\pm 0.10$ \\
Rhodes/HartRAO & WMAP K-band\tablefootmark{b} && $-2.76\pm 0.24$ & $-3.08\pm 0.13$ & $-3.32\pm 0.18$ \\
GEM &&& $-2.77\pm 0.23$ & $-3.08\pm 0.13$ & $-3.33\pm 0.20$ \\
\noalign{\vskip 2.5pt}
\hline
\noalign{\vskip 2.5pt}
\end{tabular}
\tablefoot{
\tablefoottext{a}{All-sky coverage.}
\tablefoottext{b}{Sky coverage in the GEM strip.}}
\label{Tab7}
\end{table*}

The main difference is the double-peaked distribution of the symmetric component in the temperature histogram of the Rhodes survey; whereas for the GEM survey, 
its low-emission region displays a single-peaked temperature profile similar to those in the histograms of the WMAP K-band and 408\thinspace{MHz} surveys. 
In addition, the close proximity between the lower limit of the remainder and the upper limit of the symmetric component in the histogram of the Rhodes survey
results in an exceptionally narrow transition region between the two components. This causes the considerable reduction in gray-shaded pixels that 
would otherwise mark the transition region. The GEM survey, on the other hand, shows an increase in the width of the transition region compared with
408\thinspace{MHz} that suggests an intermediate stage toward the patchy distribution seen in the WMAP K-band morphology of the transition region. A deeper 
appreciation of these implications lies beyond the scope of the present discussion, but it certainly bears relevance to the distinction that synchrotron and 
free-free mechanisms can have in the characterization of Galactic emission. An improvement in the understanding of the morphology of the large-scale structure 
of the radio continuum in high-latitude regions is a major goal of the GEM project.

\subsection{Temperature spectral index}

We close our discussion with a brief treatment of the temperature spectral index between the above set of surveys as a function of the two-component
temperature distribution. In Table \ref{Tab7} we compare the mean spectral index between the surveys listed in the first and second columns over the 
emission regions of the 408\thinspace{MHz} and WMAP K-band surveys. Between these two surveys the mean spectral index decreases from the high- to the 
low-emission regions, as expected from a Galactic latitude dependence, given the confinement of relativistic eletrons with lower energies, and therefore 
lower losses, to scale heights closer to the Galactic Plane and their sources, but also due to the flatter spectral index of HII regions. The mean spectral 
index of the high-emission region is essentially insensitive to the survey used to determine it. However, for the transition and low-emission regions, 
the mean spectral index decreases more sharply in the regions defined for the WMAP K-band survey. The same behavior can be seen when the partial-sky coverage 
in the GEM strip is used instead. 

The steepening of the synchrotron spectral index with frequency is also apparent for the GEM and Rhodes surveys. Their mean spectral index with 
respect to the 408\thinspace{MHz} is practically constant across the two emission regions and their transition. For the GEM survey, this constant behavior
is not readily verified unless the mean spectral indices are calculated with respect to the emission regions defined by the temperature distribution of 
the GEM survey itself. For this the estimates for the high-, transition-, and low-emission regions become $-2.66\pm 0.24$, $-2.56\pm 0.17$, and $-2.62\pm 0.14$, respectively. Thus the constancy of the mean spectral index between 408\thinspace{MHz} and the 2300\thinspace{MHz} can be given directly for the entire GEM 
strip, irrespective of emission region, and results in $-2.64\pm 0.11$ according to the Rhodes survey and $-2.63\pm 0.21$ according to the GEM survey. 
Guzman et al.~(\cite{Guzman11}) have also recently reported a temperature spectral index of $-2.5$ to $-2.6$ over most of the sky between 45\thinspace{MHz} and
408\thinspace{MHz}. On the other hand, when the mean spectral index of the GEM and Rhodes surveys is obtained with respect to the WMAP K-band, the steepening 
of the spectral index increases toward the low-emission region, as expected for relativistic electrons of higher energies propagating to the larger 
scale heights that characterize these regions at high Galactic latitudes. This steepening is not entirely due to synchrotron losses and has been shown 
to have a dependance on the missing subtraction of a monopole component in the lower frequency data (Kogut et al.~\cite{Kogut12}).

Deriving the spectral and spatial distribution of the synchrotron spectral index fundamentally depends on the calibration of an homogeneous baseline 
in each of the compared surveys. This dependance is very sensitive to the scanning strategy of the observations. Despite significant improvements  
applying destriping techniques, it is apparent from the results in Table \ref{Tab7} that higher dispersions are associated with mean spectral indeces 
between surveys with different scanning strategy than between surveys with a similar one. The spatial distributions of the temperature spectral index 
shown in Fig.~\ref{Fig20} have been rendered in false color shades to enhance the striping pattern of declination scans in the Rhodes and 408\thinspace{MHz} 
surveys. Stripes in declination are mainly due to differential ground pick-up through the antenna sidelobes, a systematic effect that in the GEM survey was 
significantly reduced by the choice of zenith-centered circular scans and the double-shielded configuration of the experimental set-up. 

The spectral index map between the GEM and the 408\thinspace{MHz} surveys shows a systematic drop in a rather large region in an apparent southern extension 
of the Galactic Bulge, however. A similar effect is seen in the corresponding map for the Rhodes survey over a rectangular region centered approximately on 
Cen A. Curiously, the apparent southern extension of the Galactic Bulge does not show up as an unexpected symmetrical feature in the spectral index map between 
the GEM and WMAP K-band surveys, where its distribution follows similar patterns on both sides of the Galactic Plane near the Galactic Center. This kind of 
symmetry is absent from the Rhodes and 408\thinspace{MHz} spectral index maps with respect to the WMAP K-band survey. For now, we can only expect that a
more thorough revision of the method used to calibrate the baseline of the GEM survey can possibly ressolve discrepancies of this nature. As we  
noted in Sect.~\ref{residual}, the cleaning of residual destriping revealed a tendency of the negative residuals to require larger corrections than the 
positive residuals, which could imply that the ground subtraction slightly overcorrected this type of contamination.

\begin{figure*}
\centering\includegraphics[width=14cm]{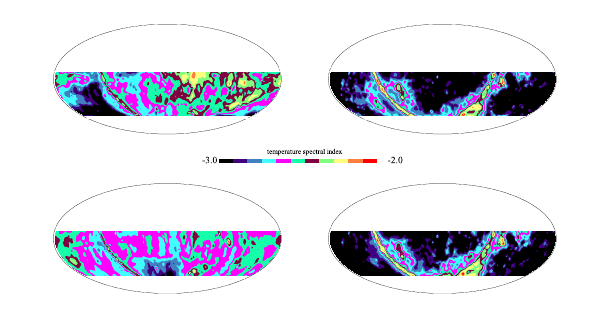}
\caption{Mollweide projection in celestial coordinates of the temperature spectral index maps for the GEM (top) and Rhodes (bottom) surveys with respect to the 408\thinspace{MHz} (left) and WMAP K-band (right)
surveys in the GEM strip.}
\label{Fig20}
\end{figure*}

\section{Conclusions and future prospects}\label{ENDE}
The preparation of low-frequency surveys of the radio continuum from the ground poses a considerable challenge to scientific endeavours, given the need to 
combine the stability of instrument performance with observational constraints over long periods of time. Any mismatch in this process introduces systematic 
effects, which radio-astronomers aim to keep under control by designing experiments that can address the desired balance between the proposed science and 
enduring logistics. The GEM project has evolved in this scenario for nearly two decades with the ultimate goal to improve our understanding of the large-scale 
structure of Galactic synchrotron emission in total intensity and polarization between 408\thinspace{MHz}\ and 10\thinspace{GHz}, signaling the ever-increasing 
impact that foreground contamination has on CMB studies and its implications for the astrophysics of the ISM.

We have presented our efforts at 2.3\thinspace{GHz}\ by combining total intensity observations obtained with a portable 5.5-m radiotelescope 
in Colombia and Brazil to produce a synchrotron-dominated template of Galactic emission with 66\% sky coverage from $\delta = -51\fdg73$ to 
$\delta = +34\fdg78$ with an average HPBW of $2\fdg31\times1\fdg85$. The main focus of our analysis has been a thorough assessment of the systematics 
that affected the observations; in particular, the ground contamination. The resulting survey shows that its sky temperature distribution into regions of low 
and high emission is consistent with the appearance of a transition region as seen in the Haslam 408\thinspace{MHz} and the WMAP K-band surveys. The 
distribution of the temperature spectral index using the GEM survey does not show a spatial dependence on these regions when calculated against the Haslam 
408\thinspace{MHz} survey; but it steepens significantly from high- to low-emission regions with respect to the WMAP K-band survey in agreement with energy losses associated with relativistic electrons and their confinement volumes within the Galaxy. Proper accounting of the synchrotron spectral steepening between 
these surveys can help in establishing useful constraints on the fractional contribution of synchrotron and free-free components in the WMAP K-band survey 
and, therefore, place upper limits on residual emission components such as spinning dust (Tello et al.,~in preparation).  

The results presented in this article summarize one of the main phases in the developing of the GEM project, which started with the total intensity
408\thinspace{MHz}\ survey of  Torres et al.~(\cite{Torres96}) and is currently dedicated to polarization and total intensity measurements of the Galactic
foreground at 5\thinspace{GHz}. The preparation of spectral index maps between the four working frequencies of the project (408\thinspace{MHz},
1465\thinspace{MHz}, 2.3\thinspace{GHz}, and 5\thinspace{GHz}) promises yet to reveal an astrophysically interesting scenario for increasing our knowledge 
and understanding of the nature and composition of Galactic emission. 

\begin{acknowledgements}
      We are enormously grateful to several generations of students and technicians at LBNL and INPE, for whom GEM was a rich learning experience. We are
particularly indebted to John Gibson, Alexandre M.R.~Alves, Luiz Arantes, and Luiz Antonio Reitano for their dedicated commitment; to Jon Aymon, Tony Banday, Justin 
Jonas, and Andrew Jaffe for helpful advice and to SLB/INPE for logistics support in Cachoeira Paulista. The destriped version of the Rhodes/HartRAO map was
reproduced courtesy of Tony Banday. The GEM project in Brazil was supported by FAPESP through grants 97/03861-2, 97/06794-4 and 00/06770-2. T.V.~acknowledges support 
from CNPq through grants 305219/2004-9, 303637/2007-2, 484378/2007-4, 308113/2010-1, 506269/2010-8. M.B.~acknowledges the support of the NATO Collaborative
Grant CRG960175. S.T.~acknowledges the support provided by Colciencias funding of the GEM project in Colombia through project 2228-05-103-96, contract 
No.~221-96. D.B.~acknowledges support from FCT -- Portugal and POCI through an SFRH/BPD grant and project grants POCTI/FNU/42263/2001 and POCI/CTE-AST/57209/2004. Last, but not least, we would like to acknowledge the referee's comments, without which the significance of this article would not
have been fully appreciated.

\end{acknowledgements}

\end{document}